# 修　士　論　文
# (Master's Thesis)

論　文　題　名(The Thesis Title)

# PARALLEL DATA READOUT FOR BACKGROUND REJECTION IN CANDLES EXPERIMENT

Date: July 29th, 2016

| | |
|---|---|
| 専 攻 名<br>(Department) | Physics |
| 氏　　名<br>(Name) | Bui Tuan Khai |

大阪大学大学院理学研究科
(Graduate School of Science OSAKA UNIVERSITY)

# CONTENT

## List of abbreviations

| | |
|---|---|
| 2νββ | Two-neutrino Double Beta Decay |
| 0νββ | Neutrino-less Double Beta Decay |
| CANDLES | CAlcium fluoride for studies of Neutrino and Dark matters by Low Energy Spectrometer |
| DAQ | Data Acquisition |
| FADC | Flash Analog-to-Digital Converter |
| ATCA | Advanced Telecommunications Computing Architecture |
| MTCA | Micro Telecommunications Computing Architecture |
| PCIe | Peripheral Component Interconnect Express |
| PC | Personal Computer |
| SpW | SpaceWire |
| GbE | Gigabit Ethernet |
| SpW-GbE | SpaceWire to Gigabit Ethernet |
| Mbps | Megabits per second |
| Gbps | Gigabits per second |

# List of figures

# List of tables

**PARALLEL DATA READOUT FOR BACKGROUND REJECTION IN CANDLES EXPERIMENT**

Bui Tuan Khai

Fundamental Nuclear Physics Group (Nomachi lab.)

## ABSTRACT

CANDLES experiment in Kamioka Underground Observatory aims to obtain the neutrino-less double beta decay (0νββ) from $^{48}$Ca. This measurement is a big challenge due to extremely rare decay rate of $^{48}$Ca. Thus, in order to obtain 0νββ, it is needed to reduce background as much as possible. Series of alpha and beta decays originated from radioactive impurities can remain as background in the energy region of 0νββ. Because they are sequential decays, we can remove them by tagging preceding and following events. This tagging method requires minimized dead-time of DAQ system. A new DAQ system was introduced in CANDLES with new FADC (Flash Analog-to-Digital Converter) modules using 8 event buffers and SpaceWire-to-GigabitEthernet network for data readout. To reduce the dead-time, we developed our DAQ system with 4 parallel reading processes. As a result, the read-time is reduced by 4 times: 40msec down to 10msec, which is in equivalent to a half of previous DAQ's read-time. With reduced read-time accompanied by multiple event buffers, the new DAQ system is realized with efficiency is very close to 100% (no event lost at 20cps, which is CANDLES trigger rate, after 63 hours of data taking). With improved performance, it is expected to achieve higher background suppression for CANDLES experiment.

# Chapter 1. INTRODUCTION

## 1. Beta decay

### ❖ Overview

Unstable nuclide tend to achieve stable states via different decay modes, and beta decay is one of these radioactive decays (alpha decay, gamma decay, neutron decay, spontaneous fission). In one nuclide, beta decay is the transition from neutron to proton or vice versa with respecting this number of nucleon exceeding limit of stability. In nuclear physics, beta decay is related to weak interaction, one of three fundamental interactions in nuclear physics (strong interaction, electromagnetic interaction and gravity interaction). There are three types of nuclear beta decays:

$$\beta^- \text{ decay:} \quad (A, Z) \rightarrow (A, Z+1) + e^- + \overline{\upsilon_e}$$

$$\beta^+ \text{ decay:} \quad (A, Z) \rightarrow (A, Z-1) + e^+ + \upsilon_e$$

$$\text{EC decay:} \quad e^- + (A, Z) \rightarrow (A, Z-1) + \upsilon_e$$

where EC stands for Electron Capture. The electron and positron emission ($\beta^-/\beta^+$ decay) are accompanied by antineutrino and neutrino, repectively. The third mode of beta decay is electron capture (EC) where nucleus absorbs one atomic electron for conversion of proton to neutron instead of emitting positron and neutrino.

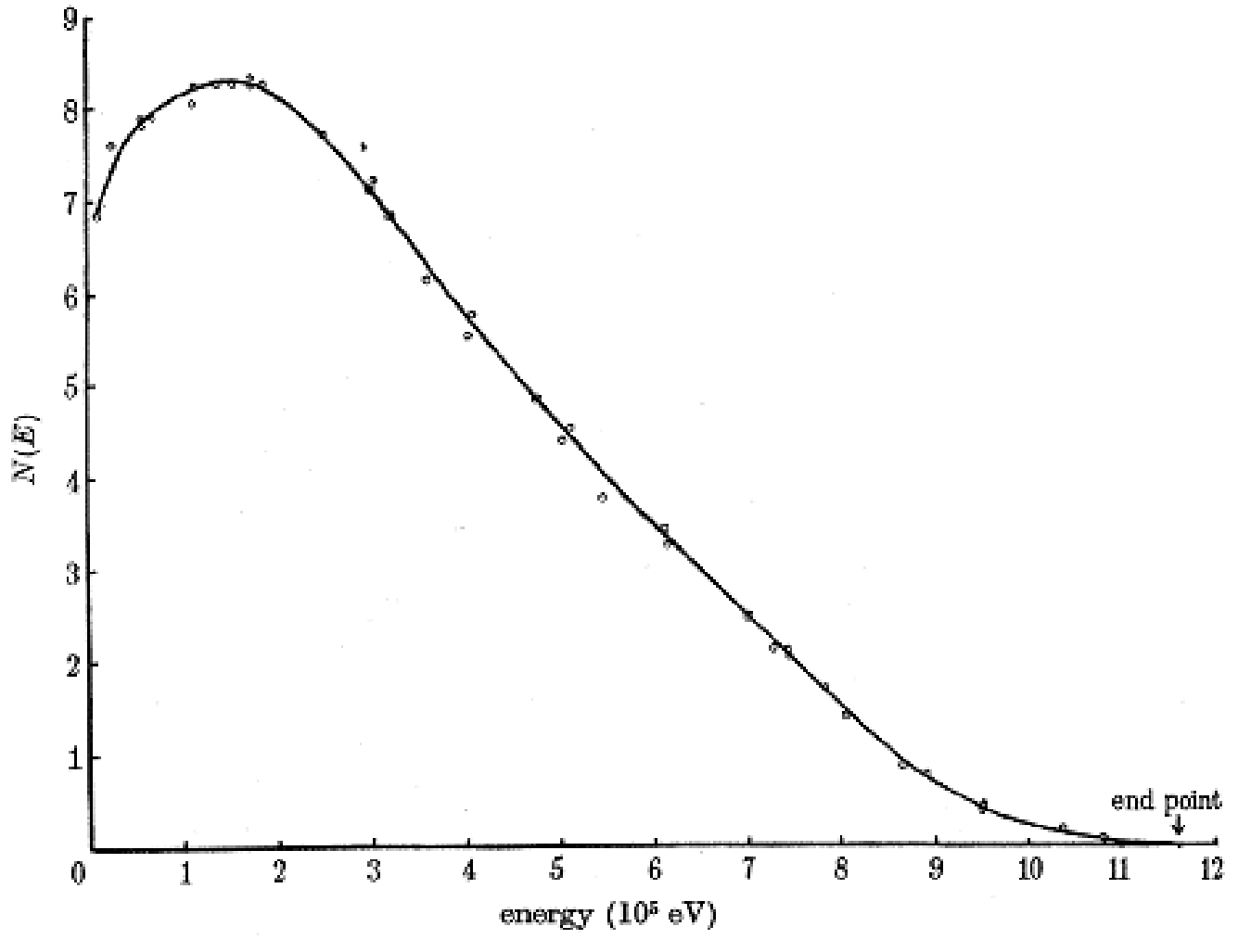

**Figure 1.1.** Energy distribution curve of $^{210}$Bi [1]

❖ **Discovery of neutrino**

Prior to the discovery of neutrinos, the following observations suggested an idea of one particle with zero-charge and negligible mass:

i. Kinetic energy of emitted electron is not mono-energy. Instead, the energy spectrum is continuous with maximum kinetic energy is $Q_\beta = M_{(A,Z)} - M_{(A,Z-1)} - M_{e^-}$. This requires a third particle in the final state of beta decay to share the released energy $Q_\beta$. Moreover, this particle cannot be massive in consideration of energy, and it has no electric charge. In the early state of beta decay studies, the beta energy spectrum was a big puzzle. Many physicist thought that was a hint of failure of energy conservation in microscopic world.

ii. Beta particle is a Fermion particle with half-integer spin. Thus, even if parent and daughter nucleus are Boson of integer spin or Fermion of half-integer spin, it is required another Fermion particle (beside electron) in the final state due to spin conservation law.

In 1932, hypothesis of a particle named "neutrino" was postulated by Pauli. This hypothesis of particle was formed to rescue the energy and spin conservation. About 20 years later, this particle was experimentally observed by C. L. Cowan and F. Reines, who got the Nobel prize in physics in 1995.

❖ **Transition laws** [2]

One can classify beta decay transition according to difference of angular momentum ($\Delta J$), isospin ($\Delta T$) and parity ($\Delta\pi$) between initial state and final state. These transitions are explained in different transition laws (or selection rules): Fermi, Gamow-Teller and forbidden transition. Additionally, factor log(*ft*), which is proportional to inversed squared nuclear matrix element, indicates transition rate in beta decays. Smaller value of log(*ft*) results in higher transition rate. Table 1.1 described various decay types with corresponding quantum numbers and log(*ft*).

- ***Fermi transition***

$$J_f = J_i \quad (\Delta J=0)$$
$$T_f = T_i \neq 0 \quad (\Delta T=0, \text{ but } T_i = 0 \rightarrow T_f = 0 \text{ forbidden})$$
$$\Delta\pi = 0$$

In Fermi transition, emitted beta particle and (anti)neutrino couple to a total spin S=0. This results in an angular momentum change $\Delta J=0$ ($\Delta L=0$). This transition law is not allowed between T=0 states.

**Table 1.1.** Selection rules and observed range of log(*ft*) values for beta decays [2]

| Decay Type | ΔJ | ΔT | Δπ | log(*ft*) |
|---|---|---|---|---|
| Supper Allowed | $0^+ \rightarrow 0^+$ | 0 | no | 3.1-3.6 |
| Allowed | 0, 1 | 0, 1 | no | 2.9-10 |
| First forbidden | 0, 1, 2 | 0, 1 | yes | 5-19 |
| Second forbidden | 1, 2, 3 | 0, 1 | no | 10-18 |
| Third forbidden | 2, 3, 4 | 0, 1 | yes | 17-22 |
| Fourth forbidden | 3, 4, 5 | 0, 1 | no | 22-24 |

- ***Gamow-Teller transition***

$$\Delta J = 0, 1 \quad \text{but } J_i = 0 \rightarrow J_f = 0 \text{ forbidden}$$
$$\Delta T = 0 \quad \text{but } T_i = 0 \rightarrow T_f = 0 \text{ forbidden}$$
$$\Delta\pi = 0$$

Gamow-Teller operator has both spin operator and isospin operator. In this transition, emitted beta particle and (anti)neutrino couple to a total spin S=1, hence, the angular momentum change $\Delta J=0, 1$ ($\Delta L=0$).

- ***Forbidden transition***

Fermi and Gamow-Teller are assumed as allowed transition ($\Delta J=0, 1$), but there is one exception of $0^+ \rightarrow 0^+$ which is super allowed transition. Other transitions with ΔJ larger than 1 are considered as forbidden transition because the transition rate of those are really small (Table 1.1).

## 2. Double beta decay (DBD)

### 2.1. Double Beta Decay

<u>Energy forbidden</u>: Double Beta Decay (DBD) is a rare nuclear weak process. It occurs due to ordinary single beta decay is energetically forbidden or large spin difference. In this section, these two reasons are discussed in more details.

This type of decay happens between even-even isobars when the decay to intermediate nucleus is energetically prohibited due to pairing interaction. This pairing interaction, which is indicated in the last term of Semi Empirical Mass Formula (Bethe and Weizsacker, 1935), shifts the binding energy of a given isobaric chain into two mass parabolas of even-even and odd-odd isobars. Beta decays take place between these isobars in order to reach lowest binding energy nucleus, which is stable. Because of pairing interaction, some decays of even-even nucleus to odd-odd nucleus is forbidden, and only double beta decay of even-even to even-even can occur. The single beta decays and double beta decays between isobaric nucleus (A=76) are described in Figure 1.2.

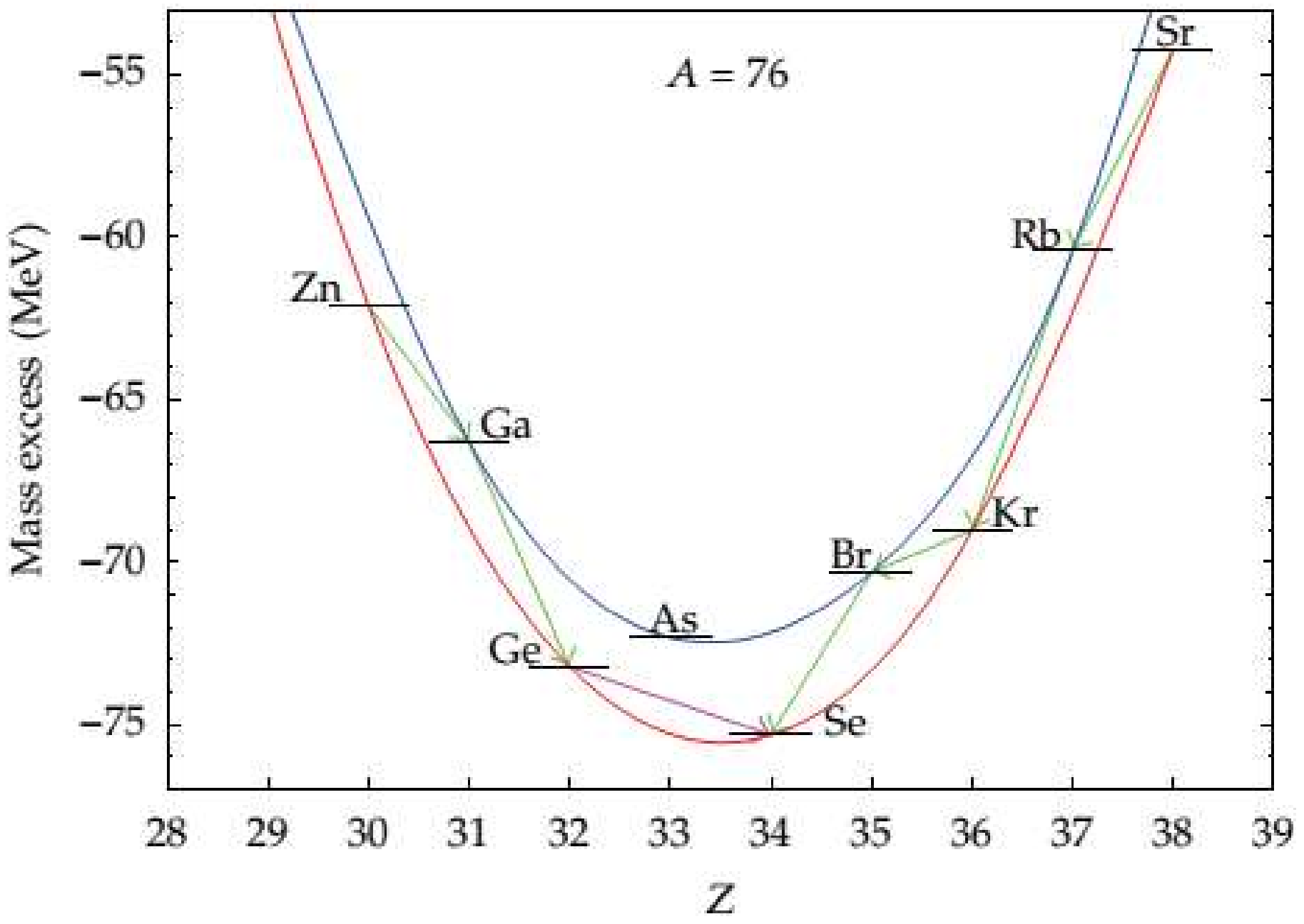

**Figure 1.2.** Binding energies as a function of atomic number (Z) of A=76 isobars [3]. Due to pairing interaction, two mass parabolas exist: even-even (N,Z) plotted in red parabola, and odd-odd (N,Z) plotted in blue parabola. Single beta decays are plotted with green arrows. There is one double beta decay from $^{76}Ge$ to $^{76}Se$, and it is plotted with green arrow.

<u>Large spin difference</u>:

Figure 1.3 describes one case of large spin difference resulting to double beta decay. Although it is not energetically forbidden, decay from state $0^+$ of $^{48}Ca$ to $^{48}Sc$'s states ($4^+$, $5^+$ and $6^+_{g.s.}$ ) is strongly suppressed due to spin transition law (forbidden transitions). On the other hand, transition from ground state of $^{48}$Ca to ground state of $^{48}$Ti is not suppressed by spin difference. From experiments, the lower limit of half-life of single beta decays of $^{48}$Ca was obtained (see Table 1.2), while half-life Two-Neutrino Double Beta Decay ($4.3\times10^{19}$ year [4]) was shorter. This proves the possibility for occurrence of Double Beta Decay is higher than single beta decay's.

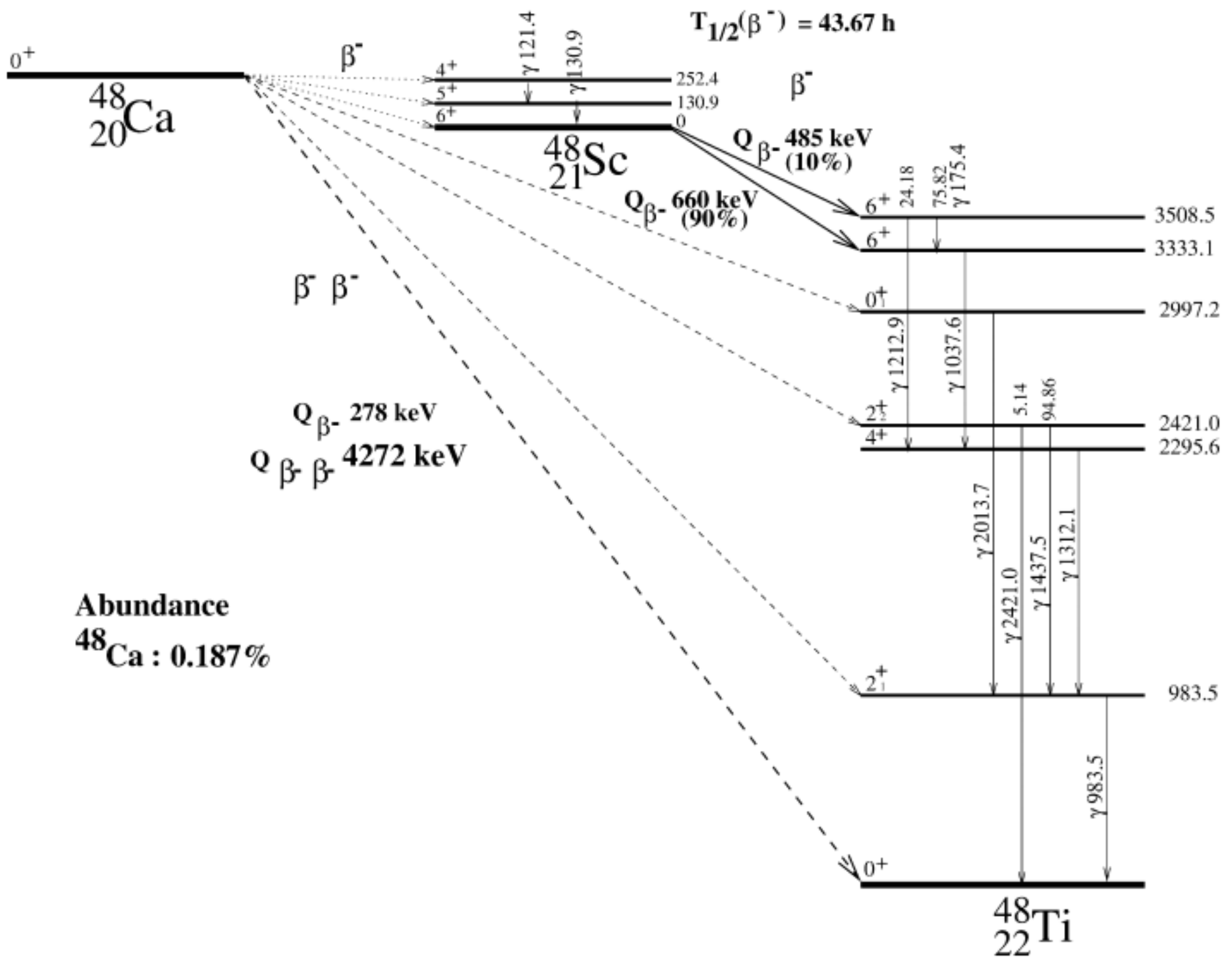

**Figure 1.3.** Decay scheme of $^{48}$Ca [5].

**Table 1.2.** Theoretical calculation and experimental result of single beta decay of $^{48}$Ca. All results are given at 90% C.L.

| **Transition** | $T^{exp}_{1/2}$ **(year)** [5] | $T^{cal}_{1/2}$ **(year)** [6] |
|---|---|---|
| $0^+ \rightarrow 6^+_{g.s.}$ | $> 0.71\times10^{20}$ | $1.5\times10^{29} \sim 1.3\times10^{31}$ |
| $0^+ \rightarrow 5^+$ | $> 1.1\times10^{20}$ | $1.1^{+0.8}_{-0.6}\times10^{21}$ |
| $0^+ \rightarrow 4^+$ | $> 0.82\times10^{20}$ | $8.8\times10^{23} \sim 5.2\times10^{28}$ |

## 2.2. Possibilities in DBD processes and experimental source

DBD can also be expressed as a conversion of nucleons (two neutrons transformed to two protons or vice versa) or quarks (two up quarks transformed to two down quarks or vice versa). Depending on number of protons and neutrons in nuclide, there are four DBD possibilities allowed in Standard Model:

$$\beta^-\beta^- \text{ decays:} \quad (A, Z) \rightarrow (A, Z+2) + 2e^- + 2\overline{\upsilon_e}$$
$$\beta^+\beta^+ \text{ decays:} \quad (A, Z) \rightarrow (A, Z-2) + 2e^+ + 2\upsilon_e$$
$$\text{ECEC decays: } 2e^- + (A, Z) \rightarrow (A, Z-2) + 2\upsilon_e$$
$$\text{EC}\beta^+ \text{ decays: } e^- + (A, Z) \rightarrow (A, Z+2) + e^+ + 2\upsilon_e$$

The released energy from these decays are distributed in lepton products and recoil nucleus, which is neglected. Respectively, these energies of decays are determined by:

$$Q_{\beta^-\beta^-} = M(A,Z) - M(A,Z+2)$$
$$Q_{\beta^+\beta^+} = M(A,Z) - M(A,Z-2) - 4m_ec^2$$
$$Q_{ECEC} = M(A,Z) - M(A,Z-2) - 2\varepsilon$$
$$Q_{EC\beta^+} = M(A,Z) - M(A,Z-2) - 2m_ec^2 - 2\varepsilon$$

where M(A,Z), M(A,Z+2) and M(A,Z-2) are atomic mass of atoms (A,Z), (A,Z+2) and (A,Z-2), respectively; ε is the excitation energy of atomic shell of daughter nucleus. As we can see, Q-value (or energy released) of $\beta^-\beta^-$ decays is higher than the others. Consequently, $\beta^-\beta^-$ decays have much higher phase-space factor, which allows higher transition probability (proportional to $Q_{2\upsilon}^{11}$ [7]). Therefore, $\beta^-\beta^-$ isotopes are more preferable in real experiments. Among 35 $\beta^-\beta^-$ isotopes, research groups consider the DBD source according to:

- Q-value: higher Q-value, we can achieve lower background.
- Natural abundance together with ease of enrichment
- Compatibility with detection technique

Q-value and natural abundance of nine preferred isotopes in DBD experiments is summarized in Table 1.3. The isotope with highest Q-value is $^{48}$Ca, next are $^{150}$Nd and $^{96}$Zr. Since the Q-value is higher than natural background (2.6MeV for γ-rays and 3MeV for β-rays), these three isotopes are easier for experiment to realize background-free condition. About the abundance, most of isotopes' abundance are in the few of percentage (%), but two typical exception: $^{48}$Ca's negative case (<0.2%) and $^{130}$Te's positive case (>34%).

**Table 1.3.** Q-value, natural abundance and half-life of 2νββ decay ( $T_{1/2}^{2\upsilon}$ ) of isotopes considered in DBD experiments. $T_{1/2}^{2\upsilon}$ values are taken from different experiments.

| Isotope | Q-value (MeV) | Abundance (%) | $T_{1/2}^{2\upsilon}$ ($10^{19}$ year) | Experiment |
|---|---|---|---|---|
| $^{48}$Ca | 4.263 | 0.187 | $4.3 \pm^{2.4}_{1.1}$ (stat.) $\pm 1.4$(sys.) | Hoover Dam [4] |
| $^{76}$Ge | 2.039 | 7.8 | $184^{+9}_{-8}$(stat.)$^{+11}_{-6}$(sys.) | GERDA [8] |
| $^{82}$Se | 2.998 | 9.2 | $9.6 \pm 0.3$(stat.) $\pm 1.0$(sys.) | NEMO-3[9] |
| $^{96}$Zr | 3.348 | 2.8 | $2.35 \pm 0.14$(stat.) $\pm 0.16$(sys.) | NEMO-3[10] |
| $^{100}$Mo | 3.035 | 9.6 | $0.711 \pm 0.002$(stat.) $\pm 0.054$(sys.) | NEMO-3[10] |
| $^{116}$Cd | 2.809 | 7.6 | $2.88 \pm 0.04$(stat.) $\pm 0.16$(sys.) | NEMO-3 [11] |
| $^{130}$Te | 2.527 | 34.08 | $70 \pm 9$(stat.) $\pm 11$(sys.) | NEMO-3 [12] |
| $^{136}$Xe | 2.459 | 8.9 | $216.5 \pm 1.6$(stat.) $\pm 5.9$(sys.) | EXO-200[13] |
| $^{150}$Nd | 3.371 | 5.6 | $0.911^{+0.025}_{-0.022}$(stat.) $\pm 0.63$(sys.) | NEMO-3[14] |

### 3. Neutrino-less Double Beta Decay

In section 1.2, we can see standard $\beta^-\beta^-$ proceeds by emitting two electrons and two anti-neutrinos. This is called Two-Neutrino Double Beta Decay: 2νββ, which is allowed in Standard Model due to no violation of quantum number conservation. This phenomenon was first observed with $^{82}$Se in 1987. Since then, it has been measured with many isotopes. Table 1.3 records the half-life of 2νββ determined by different experiments.

On the other hand, $\beta^-\beta^-$ can proceed in another way: the anti-neutrinos emitted are converted into neutrinos and then they are absorbed inside nuclide. The result is no neutrino emission. This is named Neutrino-less Double Beta Decay: 0νββ. In opposition to Two-Neutrino Double Beta Decay, Neutrino-less Double Beta Decay is not permitted in Standard Model due to the violation of lepton number conservation. The Feynman diagrams of 2νββ and 0νββ in Figure 1.4.

Neutrino-less Double Beta Decay is acquiring great interest, especially, after the confirmation of neutrino's non-zero mass demonstrated from neutrino oscillation. The interest is due to importance of this study:

**(i) Majorana or Dirac nature**

Ordinary particle composing matter has corresponding anti-particle. Particularly, charge particle has anti-particle with opposite charge. These particles are known as Dirac particles. There is another theory was suggested by Majorana. In this theory, among non-charge particles, there may exist one particle where there is no difference between it and its anti-particle. This particle is known as Majorana particle. Neutrino-less Double Beta Decay occurs only if neutrinos are massive Majorana particles. In other words, observation of this decay can prove the Majorana nature of neutrino.

**(ii) Violation of lepton number conservation**

There is no neutrino emitted in 0νββ decay. Thus, difference of lepton number before and after the decay is not conserved (ΔL = 2). This lepton number non-conservation is not allowed in Standard Model and we need a new physics beyond it.

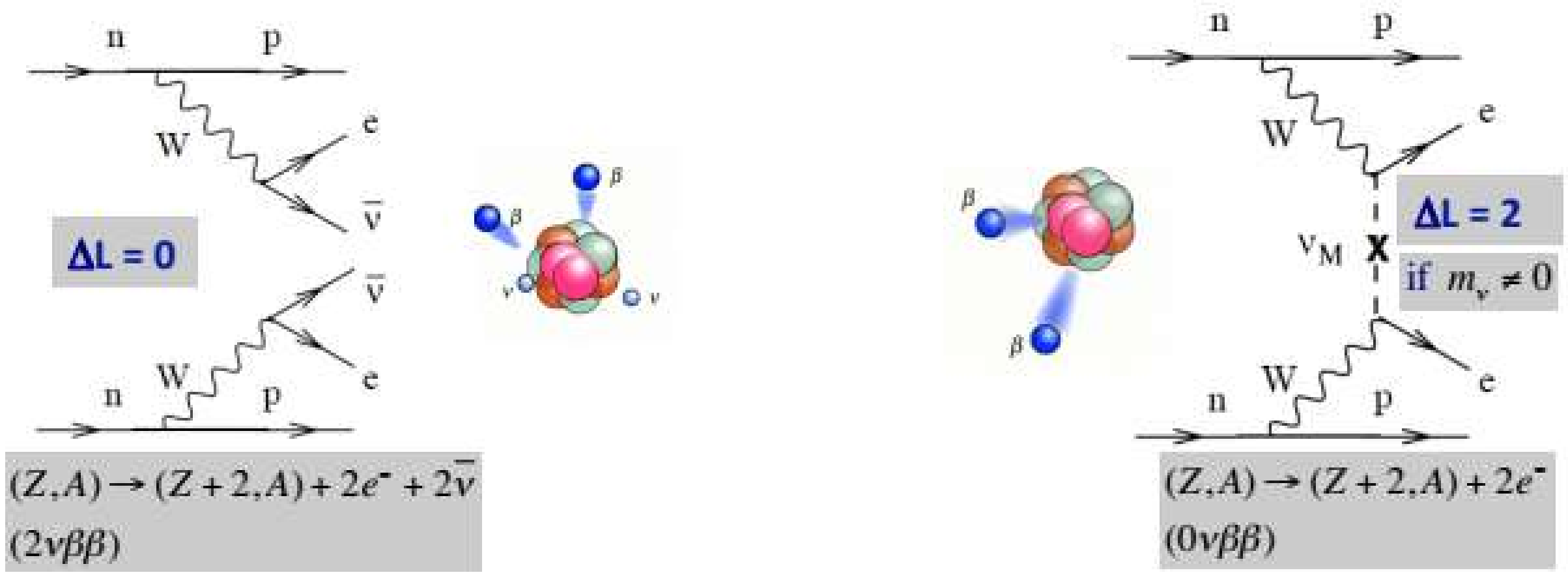

**Figure 1.4.** Feynman diagrams of 2νββ (left) and 0νββ (right)

### (iii) Neutrino mass and mass hierarchy

If half-life of 0νββ decay is practically obtained, it is possible to deduce the absolute value of neutrino mass based on a formula describing the relationship between half-life and neutrino mass. The observations of neutrino oscillation [5, 6] prove that neutrinos have non-vanish mass and mass differences are confirmed. However, absolute mass scale has not been measured. Additionally, if neutrino mass is determined, it is possible to identify the neutrino mass hierarchy (Normal Hierarchy, Inverted Hierarchy or Quasi-Degenerated).

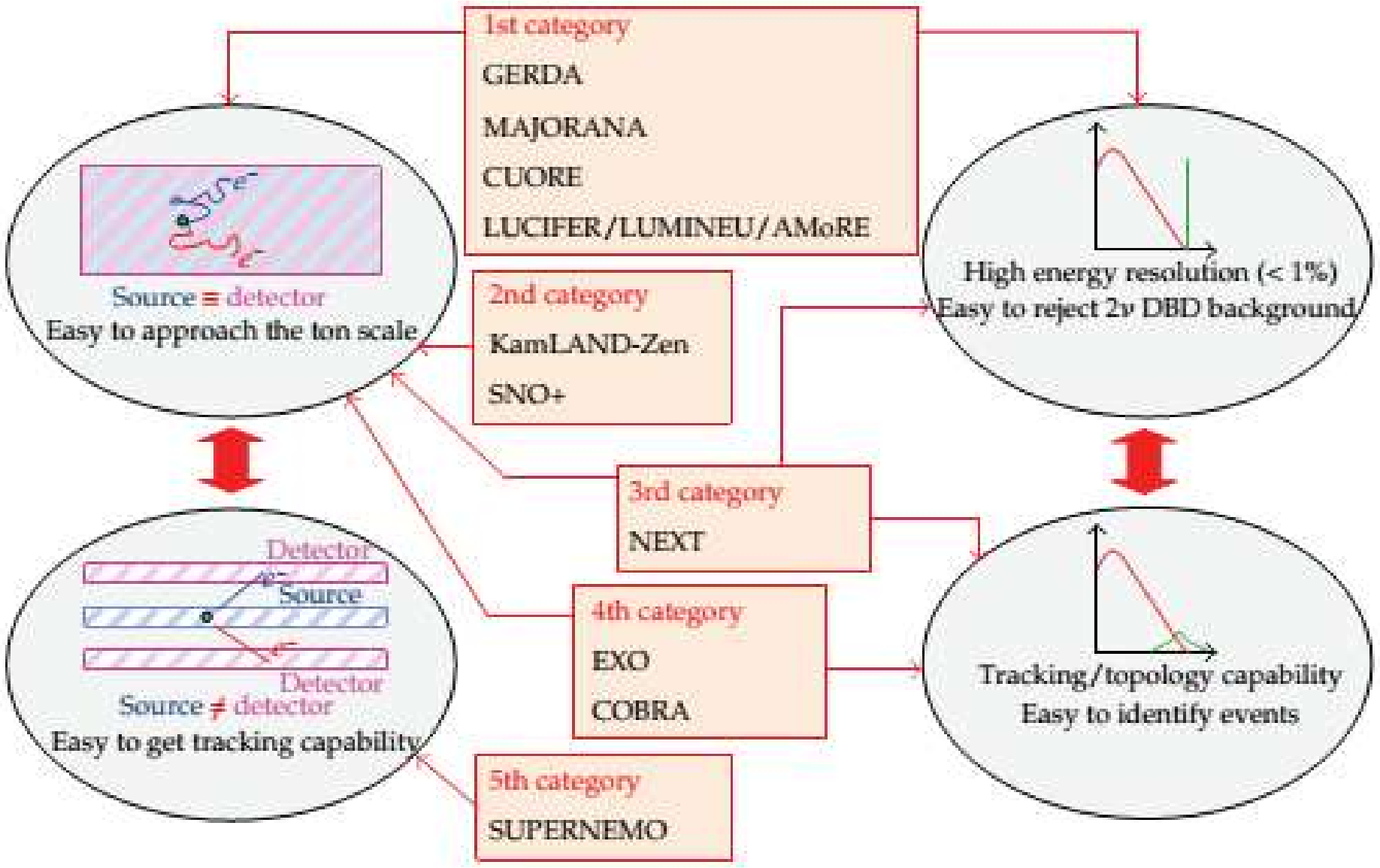

**Figure 1.5.** Experiments reviewed in the text are divided into five categories, according to the experimental approach and the main features of the detector performance [15].

## 4. 0νββ experiments

There are several experiments constructed underground aim to obtain this 0νββ by measuring different $\beta^-\beta^-$ isotopes. They can be categorized by the configuration of radiation source and detector: source and detector are the same (source≡detector), source and detector are different (source≠detector). The 0νββ experiments can be divided in to 5 categories [15]. Figure 1.5 is a summary of these categories. Different techniques are being developed to approach the 0νββ observation. So far, there has been no experimental result of absolute 0νββ half-life or neutrino mass. Several limit of $T_{1/2}^{0\upsilon}$ and neutrino mass are indicated in Table 1.4.

**Table 1.4.** Limits of 0νββ half-life $T_{1/2}^{0\upsilon}$ and neutrino mass $<m_{\beta\beta}>$ of several isotopes. All limits are at 90% C.L.

| **Isotopes** | $\mathbf{T_{1/2}^{0\upsilon}}$ **(year)** | $\mathbf{<m_{\beta\beta}>}$ **limit (eV)** | **Exposure (kg-yr)** | **Experiment** |
|---|---|---|---|---|
| $^{48}Ca$ | $> 5.8\times10^{22}$ | 3.5 – 22 | 0.015 | ELEGANT VI [16] |
| $^{76}Ge$ | $> 2.1\times10^{25}$ | 0.24 – 0.48 | 16.4 | GERDA [17] |
| $^{82}Se$ | $> 2.1\times10^{23}$ | 1.0 – 2.8 | 4.9 | NEMO-3 [18] |
| $^{96}Zr$ | $> 9.2\times10^{21}$ | 7.2 – 19.5 | 0.031 | NEMO-3 [8] |
| $^{100}Mo$ | $> 1.1\times10^{24}$ | 0.3 – 0.9 | 34.7 | NEMO-3 [19] |
| $^{116}Cd$ | $> 1.7\times10^{21}$ | 1.4 – 2.8 | 0.14 | Solotvina [19] |
| $^{130}Te$ | $> 2.8\times10^{24}$ | 0.3 – 0.7 | 19.75 | CUORICINO [21] |
| $^{136}Xe$ | $> 1.9\times10^{25}$ | 0.16 – 0.33 | 89.5 | KamLAND-Zen [22] |
| $^{150}Nd$ | $> 1.8\times10^{22}$ | 4.0 – 6.3 | 0.093 | NEMO-3 [13] |

# Chapter 2. CANDLES experiment

In this chapter, set up of CANDLES detector is described in more details. Additionally, since low background condition is required in our experiment, sources of backgrounds are listed with associated background shielding or suppression. Because of the relation of background study and my research, research motivation is also mentioned.

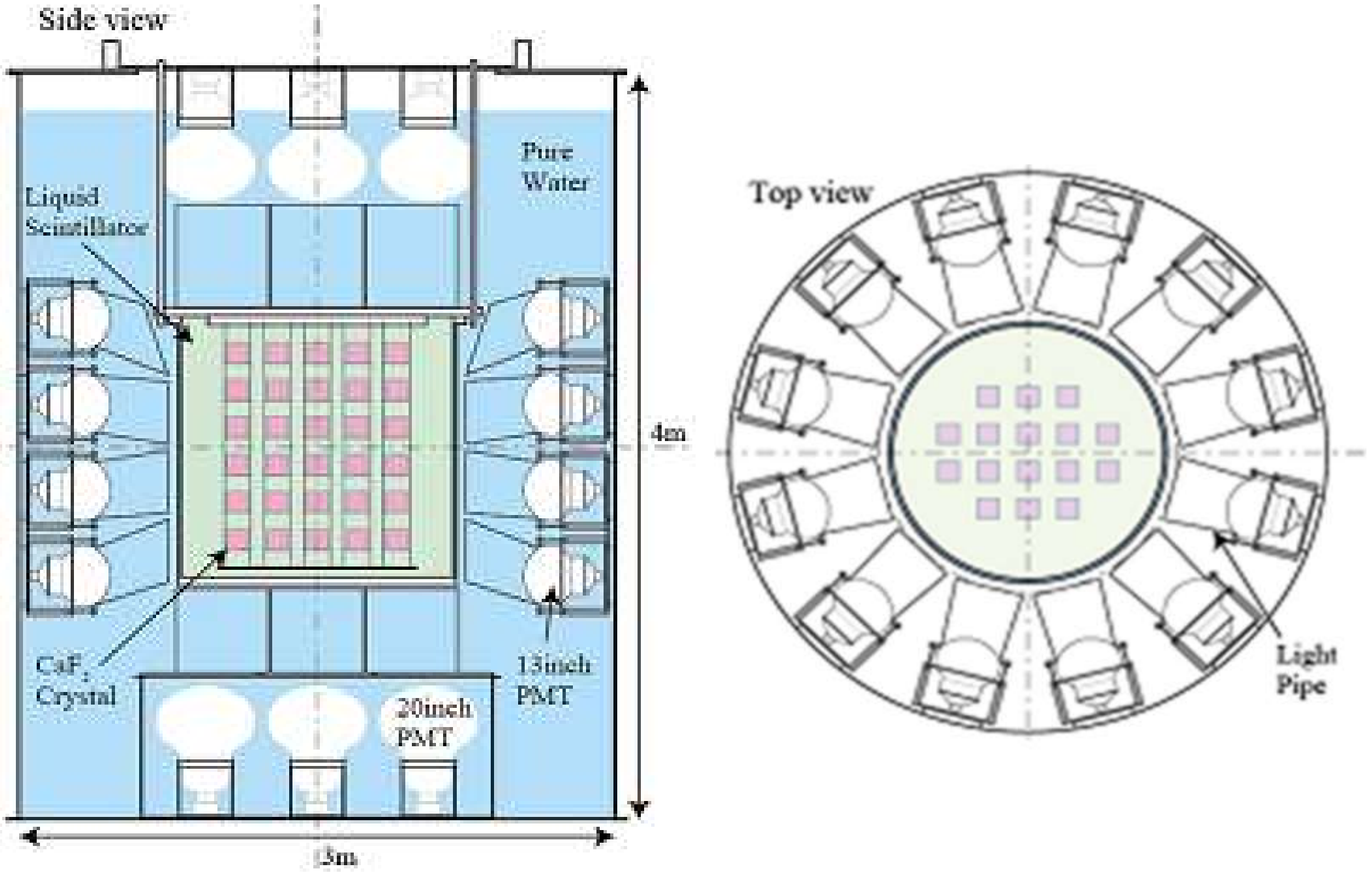

**Figure 2.1.** Arrangement of CANDLES experiment. Side view and top view are plotted on left-handed side and right-handed side, respectively.

## 1. CANDLES experiment

### 1.1. Set up

CANDLES (CAlcium fluoride for studies of Neutrino and Dark matters by Low Energy Spectrometer) is the successor of ELEGANT (ELEctron GAmma-ray Neutrino Telescope) experiment. CANDLES is constructed in Kamioka Observatory, whose vertical depth is 2.05±0.15 km.w.e. It aims to obtain $0\nu\beta\beta$ from $^{48}Ca$ by using $CaF_2$ crystals as detectors and also the source.

In CANDLES, we set up our experiment as source≡detector category. Pure $CaF_2$ crystals, which are scintillator crystals, are used. $^{48}Ca$ isotopes are obviously contained inside these crystals, this means we have 4π detection geometry. Figure 2.1 describes the arrangement of detectors including crystals, liquid scintillator, Photo-Multiplier Tubes (PMTs) and pure water. There are 96 cubic crystals (10×10×10 cm) mounted in one jig. This jig has 6 layers of $CaF_2$ crystals, and 16 columns in total are used. Total mass of these 96 crystals is nearly 300 kg (density of $CaF_2$ is 3.18 g/cm$^3$ [23]), hence, mass of $^{48}Ca$ is about 300g (natural abundance is 0.187 %, as showed in Chapter 1). These crystals are put in a liquid scintillator (LS) vessel filled with 2000 liter of LS. LS is used for 4π active shielding, which is discussed with more details in section 2. Scintillation photons emitted from $CaF_2$ and liquid scintillator are obtained by 62 PMTs surrounding. These PMTs consist of 48 PMTs with 13 inches of diameter on the side and 14 PMTs with 20 inches of diameter at top and bottom. Everything is mounted inside a cylindrical water tank (3 meter of diameter and 4 meter of height) which is made of stainless steel and filled with pure water (28000 liter). Details of geometry set up in CANDLES is referred to [24].

### 1.2. Scintillator

#### ❖ Pure $CaF_2$ crystals

In previous generation, the ELEGANT VI, $CaF_2$(Eu) crystals were used. After the latest value of $^{48}Ca$ 0νββ half-life obtained from ELEGANT VI, it was decided to increase amount of $^{48}Ca$ by three dimensional expansion in order to achieve better sensitivity. The short attenuation length of $CaF_2$(Eu), which is about 10cm [25], is really a big problem to expand the size to the order of meters. On the other hand, pure $CaF_2$ has long attenuation length, which is about 10m. According to this advantage, crystals with high purity have been developed. Scintillation light is not attenuated even the size is large. This allows to acquire accurate energy information. Due to high purity, it is possible to produce crystal with small amount of radioactive impurities. However, the $CaF_2$ has short wavelength (UV region) below sensitive wavelength of PMT used in CANDLES, and photon production of pure $CaF_2$ is almost half of $CaF_2$(Eu) (Figure 2.2 and Figure 2.3a). To overcome this, a Wavelength Shifter (WLS) coating outside $CaF_2$ is used.

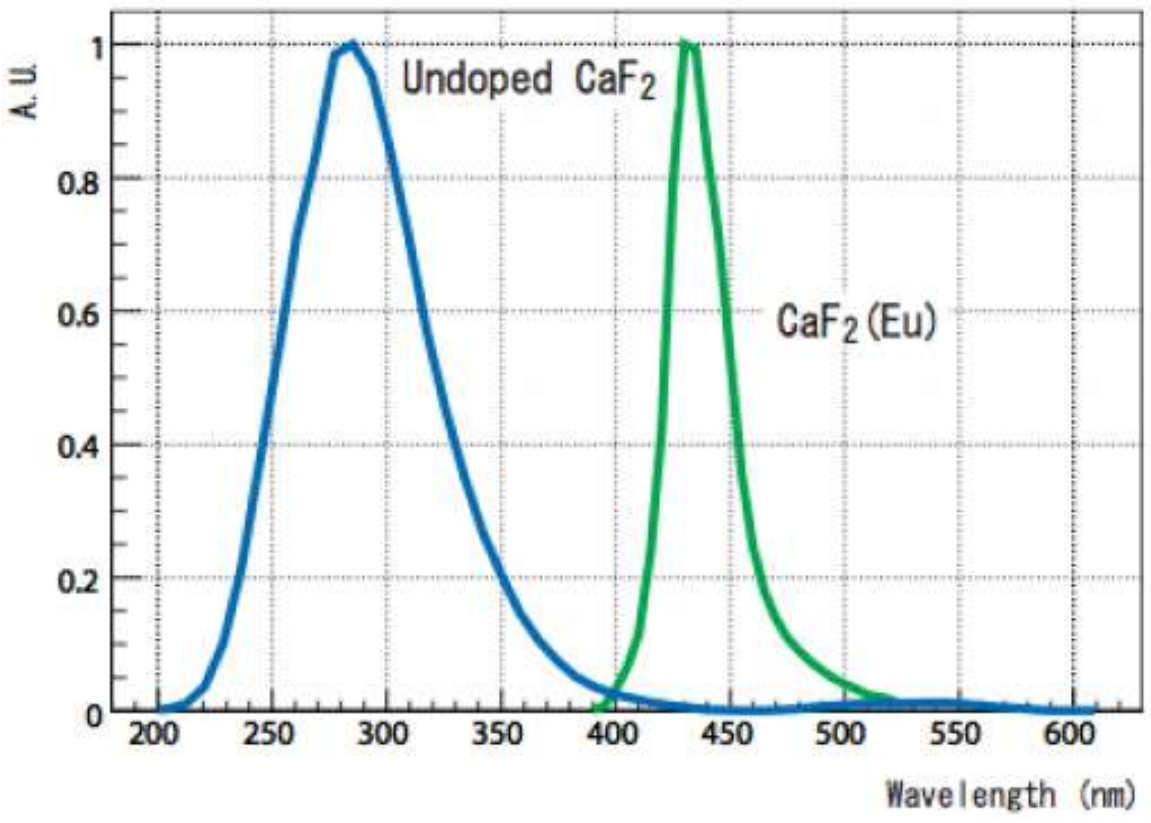

**Figure 2.2.** Emission wavelength of pure $CaF_2$ is plotted in blue and $CaF_2$(Eu)'s is plotted in green. Histogram is taken from [25]

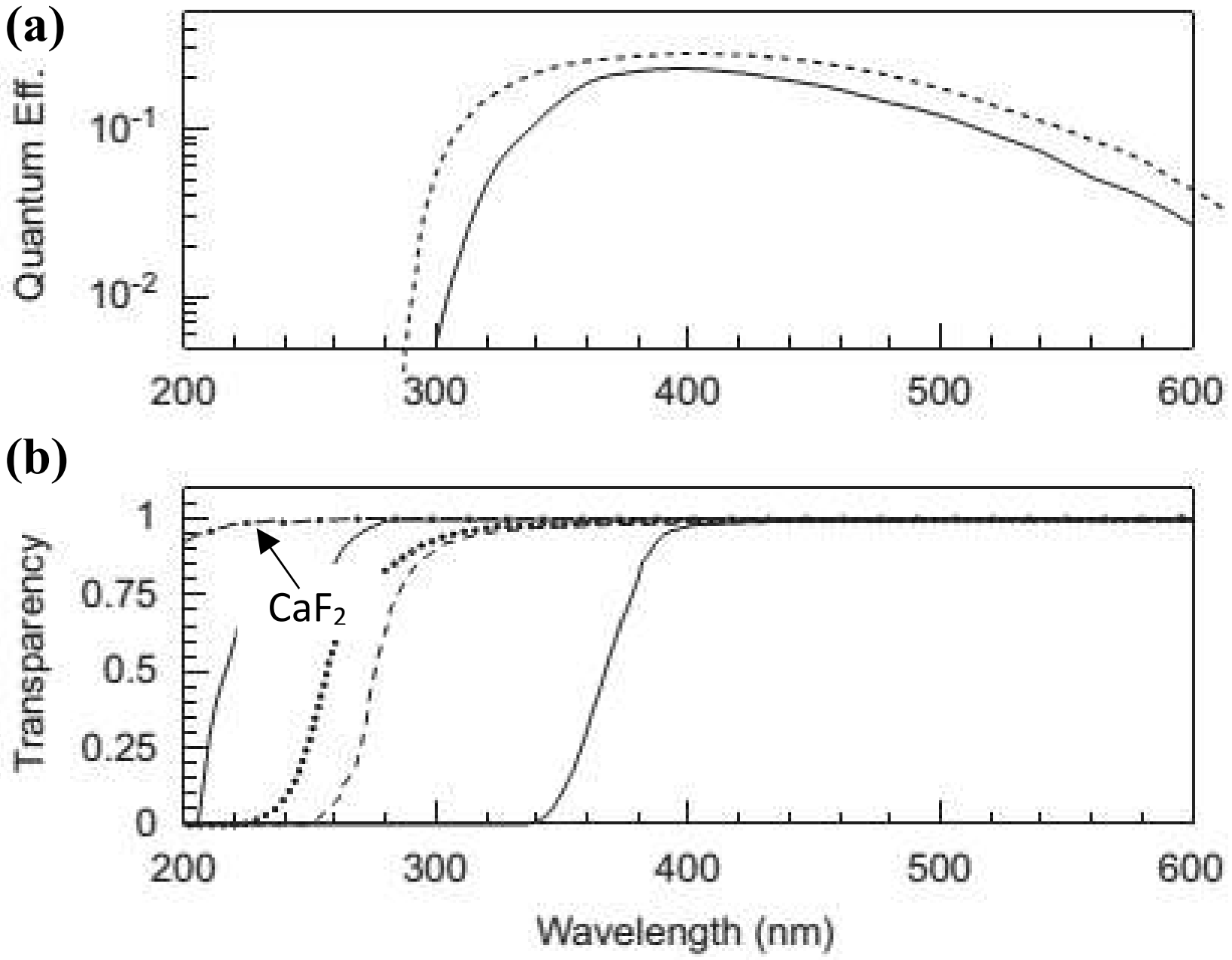

**Figure 2.3.** Some properties of optical components in CANDLES [26]. (a) Quantum efficiency of 13" PMT in CANDLES (solid) and commercial 2" PMT (dashed). (b) Transparency of $CaF_2$ plotted with dashed-dot line (and other optical components used in CANDLES: LS, acrylic resin, etc.).

❖ **Wavelength Shifter (WLS)**

WLS dissolved in liquid scintillator (LS) solvent works for scintillation light of $CaF_2$ crystals and LS solvent. UV light of $CaF_2$ propagate into LS without absorption due to the transparency of $CaF_2$ with its own emitted wavelength. WLS shifts the UV lights from $CaF_2$ to visible-region light region (about 420nm), where we can achieve nearly highest quantum efficiency. Wavelength of absorbed light and shifted light are about 360nm and 420nm (Figure 2.4). Most optical components are transparent to the wavelength about 400nm, as we can see in Figure 2.3b. With shifted wavelength, it is possible to scale up the detector to achieve the size in the order of meter.

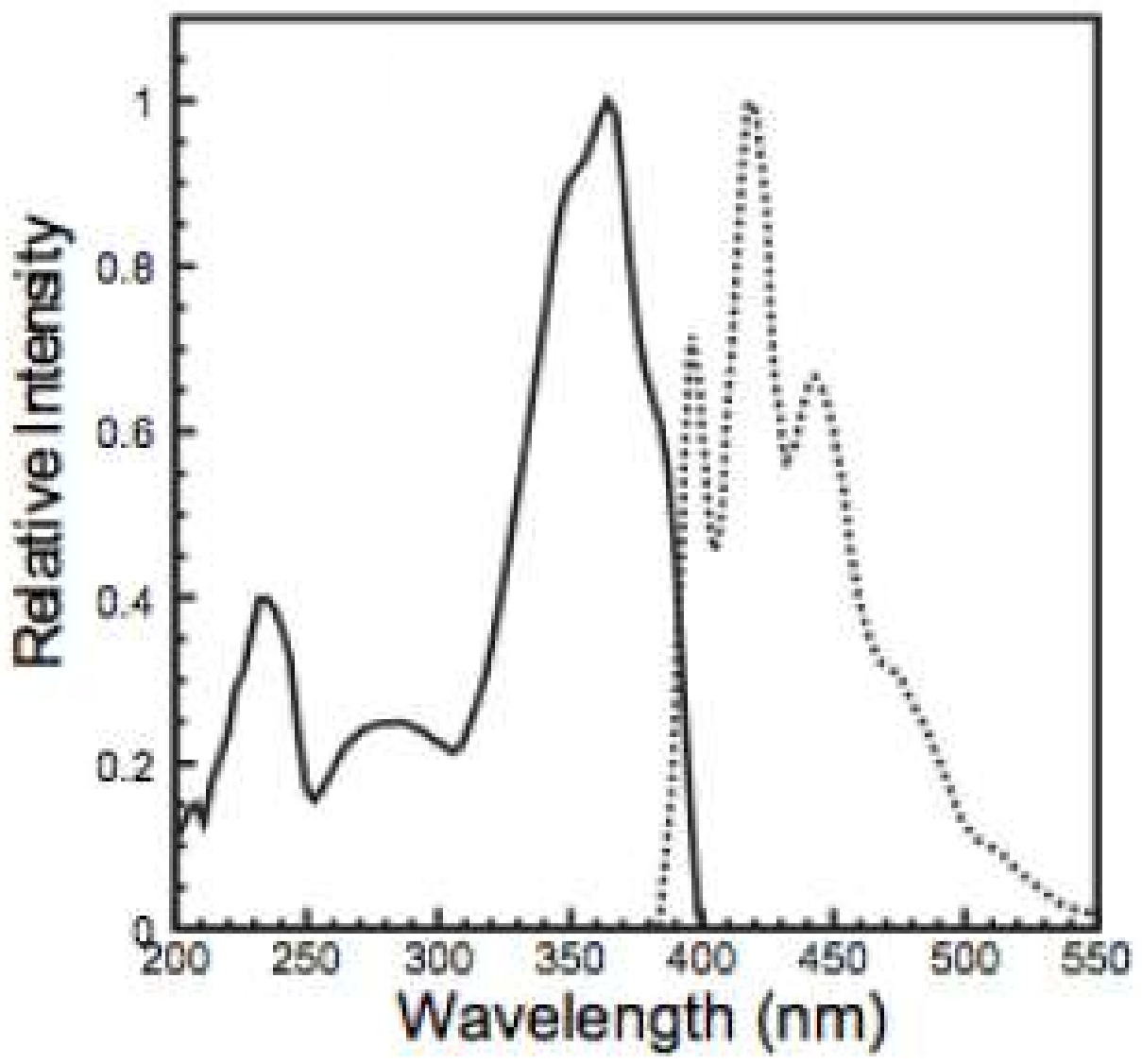

**Figure 2.4.** Wavelength spectra of the solute Bis-MSB in WLS. Wavelength of absorbed light (solid line) and shifted light (dotted line) are 360nm and 420nm, respectively [25].

❖ **Liquid Scintillator (LS)**

LS acts as an active veto. If $0\nu\beta\beta$ event occurs, scintillation light would be produced inside the $CaF_2$ crystal where it occurs, not LS. Thus, photons generated from LS are generated from unexpected radiations, which can be gamma rays (from external source or from internal decays) or cosmic rays. The decay constant of pulses from $CaF_2$ and LS are very different: $\tau_{CaF2}$ is about 1$\mu$sec and $\tau_{LS}$ is about 10nsec. According to the big difference of pulse shape, we can discriminate signals between $CaF_2$'s and LS's. Figure 2.5 is an illustration of waveforms from scintillators in three different cases: signal from LS, signal from LS and $CaF_2$ and signal from $CaF_2$. Besides that, LS is required to have large light output, high transparency and economical price due to large amount using.

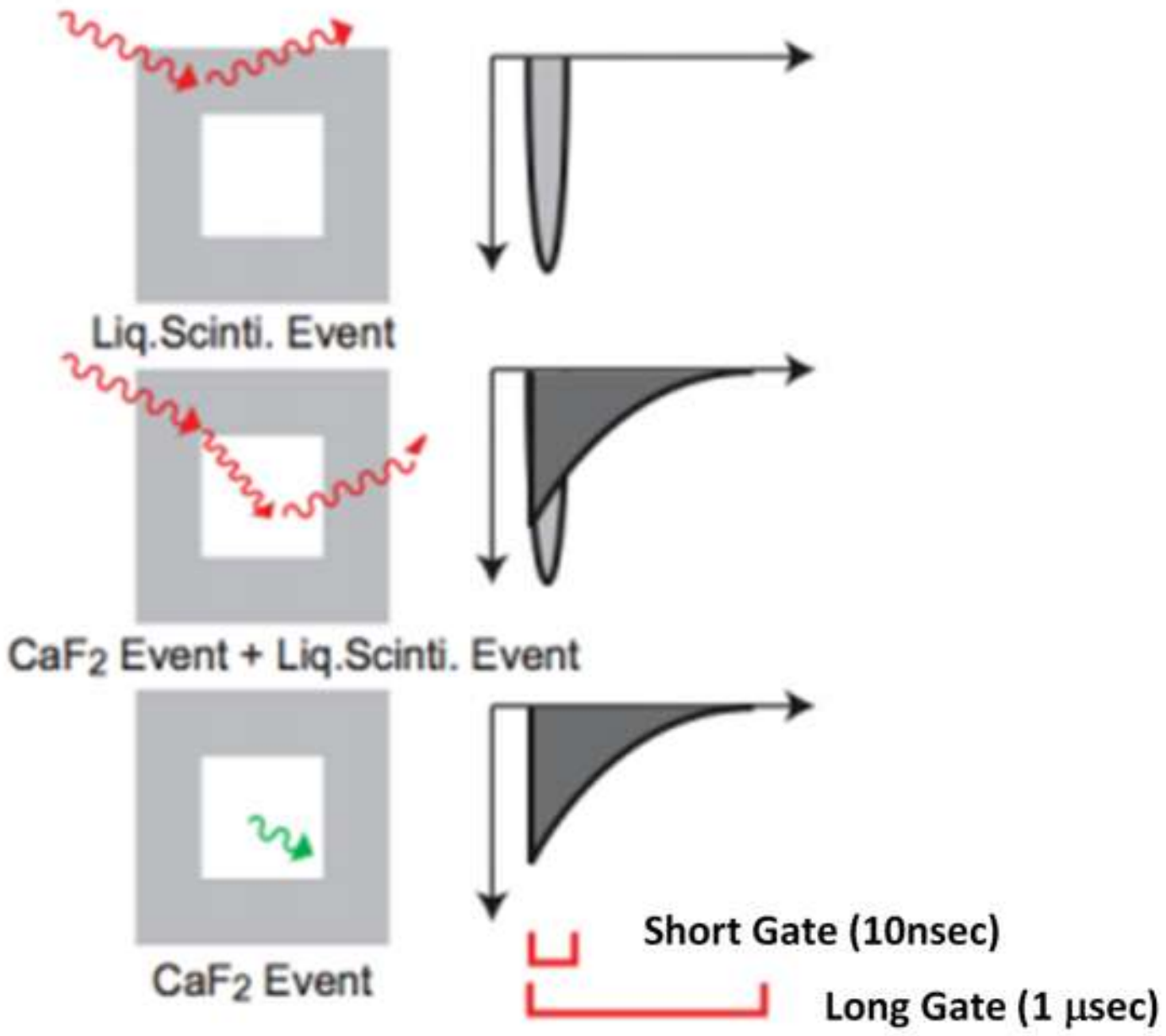

**Figure 2.5.** Illustration of waveforms from scintillators (extracted from [25]). The decay constant of pulses from $CaF_2$ and LS are about 1μsec and 10nsec, respectively.

## 1.3. Photo-Multiplier Tube (PMT)

### ❖ PMTs

The Hamamatsu R8055MOD PMTs with 13 inches in diameter are used in CANDLES. Quantum Efficiency of this kind of PMT is plotted in Figure 2.3 as a function of wavelength. Since UV sensitive PMT is not available, WLS is chosen to be well matched the sensitive region. Since the light output from the pseudo crystal (assembled with $CaF_2$ crystals and WLS) is not most sensitive region of PMT, scintillation photons would be reduced. For high light collection efficiency, PMTs are arranged surrounding the vessel which includes $CaF_2$ crystals and LS. The high light collection efficiency leads to large photoelectrons collection. Large photoelectrons collection would compensate the disadvantage of emitted wavelength from pseudo crystals.

❖ **Pure water buffer**

Pure water buffer is used as a passive shielding, which is employed between PMTs and LS vessel. It used to shield most of neutron-induced backgrounds from PMTs or external source.

## 2. Background in CANDLES

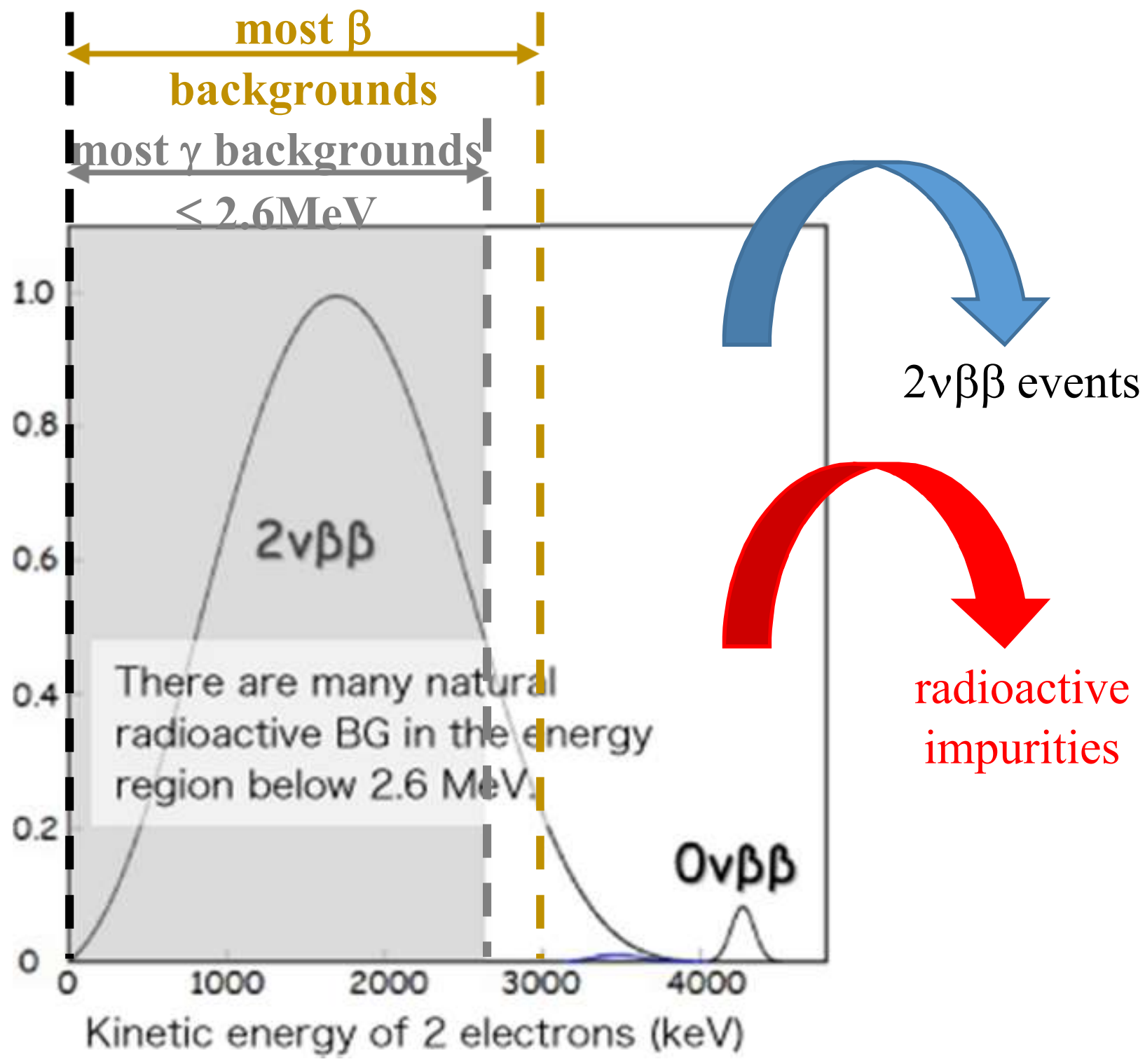

**Figure 2.6.** Electron spectrum of double beta decays of $^{48}Ca$ [27]. Possible backgrounds whose energies higher than 3MeV may come from: radioactive impurities and 2νββ.

The big advantage $^{48}Ca$ is its high Q-value: about 4.27MeV (highest in $\beta^-\beta^-$ isotopes). Considering most natural activities, most backgrounds from γ-rays is up to 2.6MeV and most of backgrounds from β-rays is up to 3MeV. Q-value of $^{48}Ca$ is far from most of natural activities, thus, it is easy to reach the background-free condition. However, in the region higher than 3 MeV, there are still backgrounds affecting 0νββ observation. These backgrounds are described in this section.

## 2.1. Two-Neutrino DDB – irreducible background

In experiment point of view, $0\nu\beta\beta$ events and $2\nu\beta\beta$ events are exactly the same. The difference is energy (Q-value). Since the Q-value of these two events are very close, it is required to have a good energy resolution at Q-value region. We use $CaF_2$ which is a scintillation crystal. Number of observed scintillation photons influence the resolution of scintillation detector. The good thing of CANDLES is $4\pi$ coverage (PMTs surrounding $CaF_2$) and its transparency (attenuation of light before reaching PMT is negligibly small). Thus, to increase the resolution, we have development of light collecting system with light pipes applied to all PMTs. Moreover, temperature also affects the resolution of scintillation detectors [28]. Therefore, we also stabilize temperature in $CaF_2$ crystals. Details of development and performance can be seen in [29].

## 2.2. Gamma-rays from (n,γ) reactions

Liquid Scintillator is strong active veto. Nevertheless, unexpected events produced by high energy γ-rays from (n,γ) reactions can also contribute in CANDLES background at the interesting region (4.27MeV). According to data analysis and simulation, these are the most dominant background in CANDLES. Specific γ peaks observed are 7.5 MeV and 9 MeV. Thermal neutrons are captured in rich material such as stainless steel, used in water tank of CANDLES, or rock, γ-rays are emitted and deposit their energy in $CaF_2$ crystals.

Method for rejection: A passive shielding for neutron was constructed in 2015. It consist of Si rubber sheet containing of $B_4C$ inside and outside the detector and a lead layer with 10 cm to 12 cm of thickness. The shielding design is optimized from simulation and it is expected to reduced (n,γ) background down to 1/80 level of the current status.

## 2.3. Background from impurities

Impurities exist inside CANDLES detector can contribute as background around Q-value region. There are 2 kinds of impurity background that we have to consider as background in the region of Q-value of $^{48}Ca$ (4.27MeV):

**a. Consecutive events (BiPo decays)**

Figure 2.7 is a pulse shape of consecutive events in CANDLES. These events are originated from sequential decays in natural radiations and they meet following conditions:

- Half-life of daughter nuclide is short and, hence, it can influence high probability of short event interval. The decay constant of $CaF_2$ is about 1μsec and window of FADC is about 8μsec. Thus, we can observe these consecutive events in pile-up pulse shape.
- Total energy obtained by $CaF_2$ crystals in unit of MeV electron equivalent (MeVee) can influence on Q-value region of $^{48}Ca$.

With above requirements, we can find two sequential decays that can behave as background: decays from $^{214}Bi$ (β-decay) and $^{214}Po$ (α-decay) in Uranium-series, and decays from $^{212}Bi$ (β-decay) and $^{212}Po$ (α-decay) in Thorium-series.

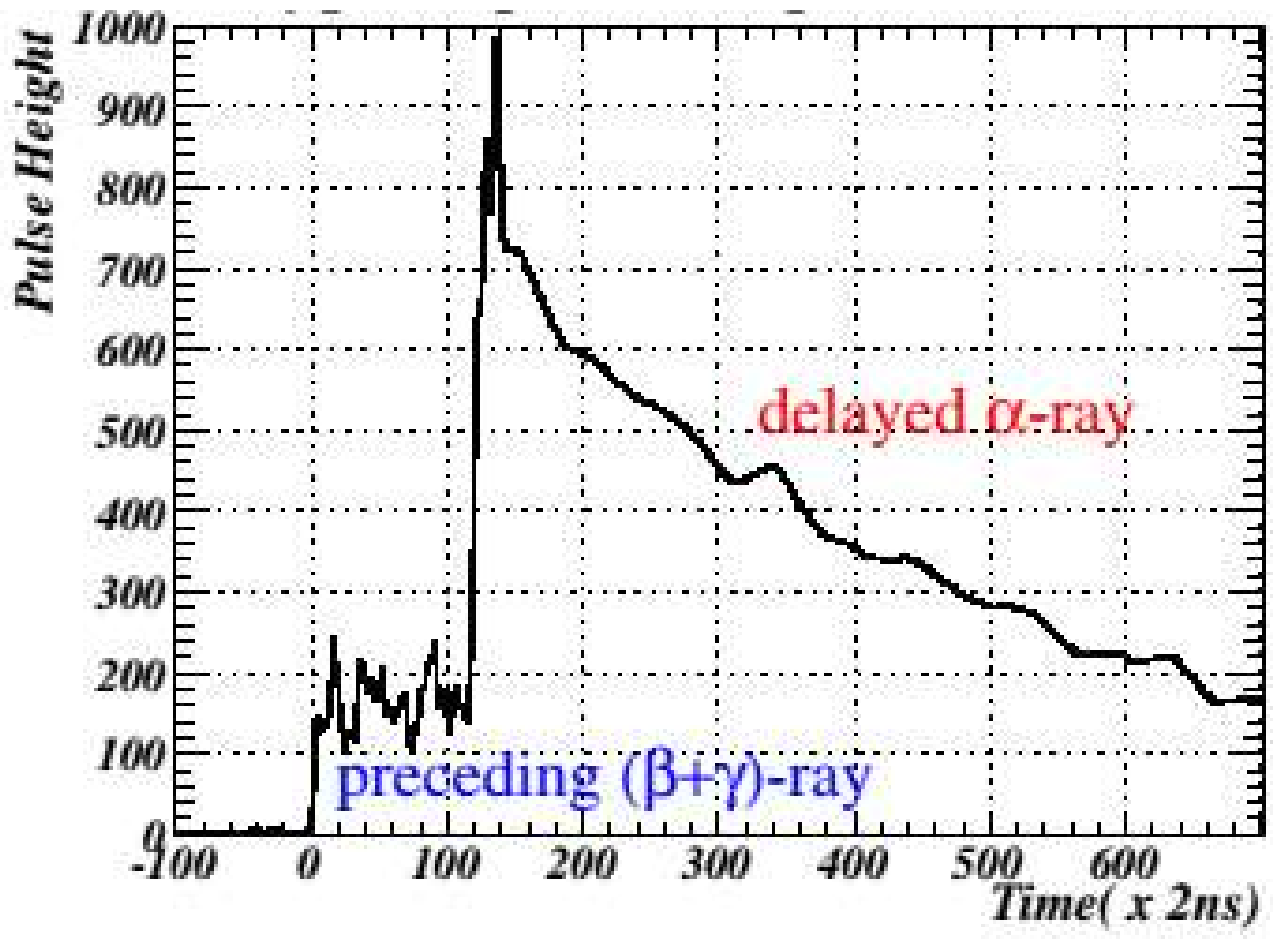

**Figure 2.7.** Pulse-shape of a consecutive event (β-decay then α-decay) behaving as background in CANDLES

Figure 2.8a and 2.8b inform details of released energy and half-life of these decays. Due to the different of particle type (α and β particles), lights output generated by α-particle is adjusted by quenching-factor. In CANDLES, by using $CaF_2$ crystals as scintillator detector, quenching factor of these α-decays (from $^{214}Po$ and $^{212}Po$) is about 35% [30]. With quenching factor, total energy of two consecutive events $^{214}Bi$-$^{214}Po$ and $^{212}Bi$-$^{212}Po$ are 5.8MeV and 5.3MeV, respectively [30]. As we see, these consecutive events are only originated from Bi and Po, so they are also named BiPo decays.

Method for rejection: These BiPo decays have typical pulse-shapes, so we apply PSD (Pulse Shape Discrimination) to identify these backgrounds. Currently, this kind of background is reduced 3 times of magnitude [30].

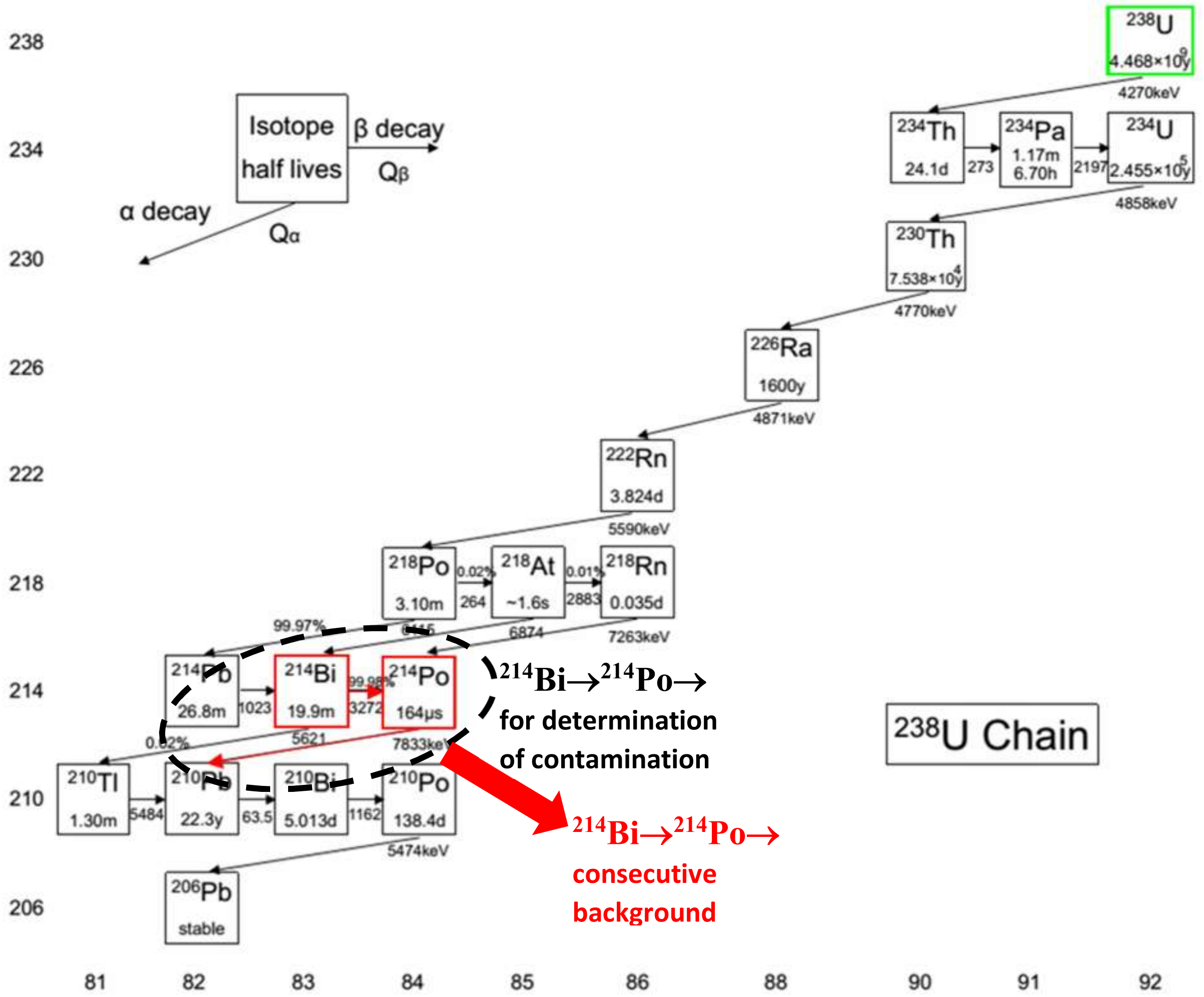

**Figure 2.8a.** Serious background in CANDLES from U-series as pile up event is marked in red color. Two consecutive decays of beta from $^{214}$Bi and alpha from $^{214}$Po forms a pile-up event in FADC window since the half-life of $^{214}$Po is short (164 μsec). Current decays used for determination of contamination ($^{214}$Bi→$^{214}$Po→) are marked with black circle.

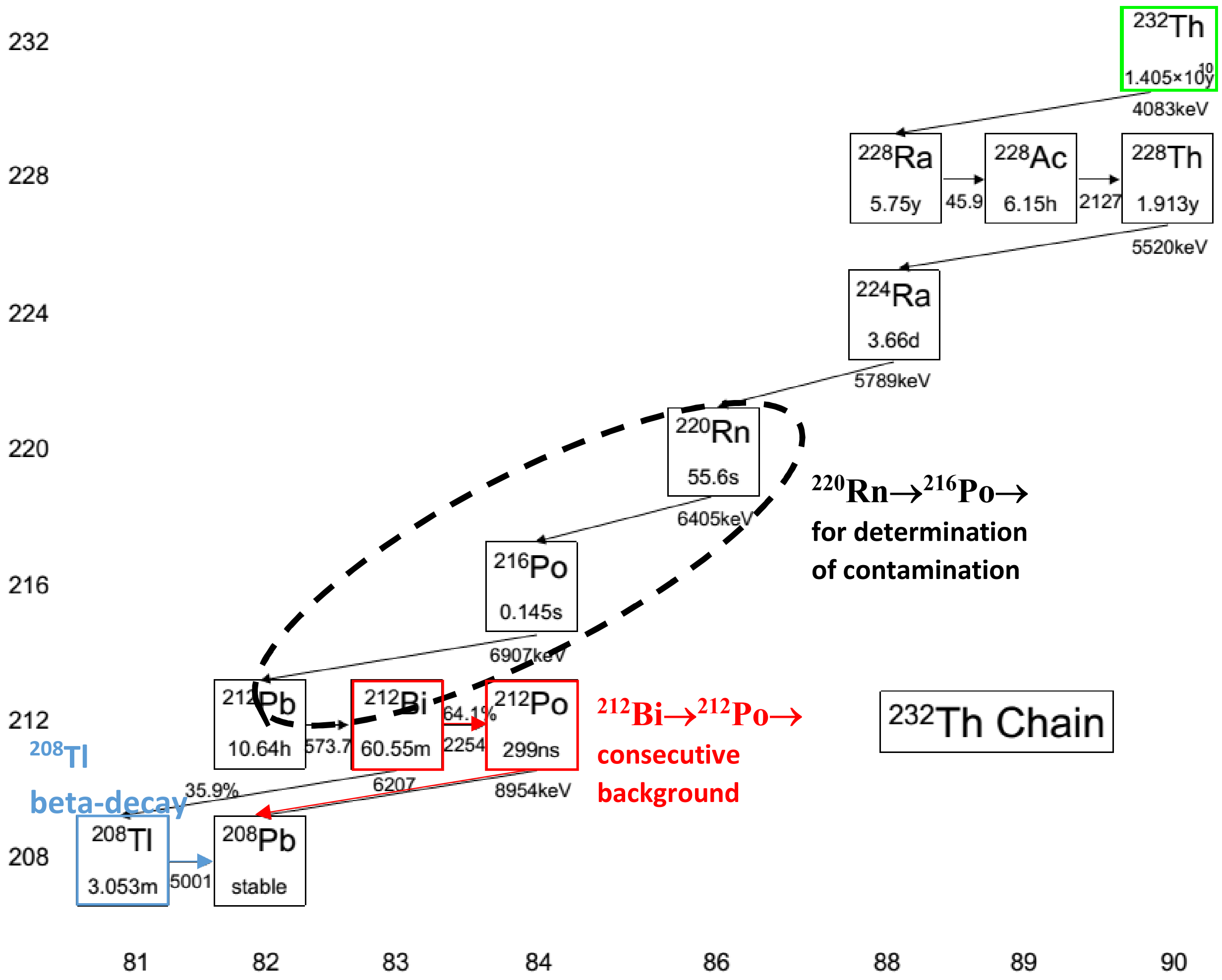

**Figure 2.8b.** Serious backgrounds in CANDLES from Th-series are marked in color (pile-up event is marked in red color and beta-decay event is marked in blue color). Two consecutive decays of beta from $^{212}$Bi and alpha from $^{212}$Po forms a pile-up event in FADC window since the half-life of $^{212}$Po is short (299 nsec). Current decays used for determination of contamination ($^{220}$Rn→$^{216}$Po→) are marked with black circle.

### b. Backgrounds from β-decays of $^{208}$Tl

Among β-decays in natural radioactivity, there are two β-decays having high $Q_\beta$ and effect to interested region $Q_{\beta\beta}$ (Q-value) of $^{48}$Ca: β-decay of $^{210}$Tl in Uranium-series with $Q_\beta$ is 5.484MeV, and $^{208}$Tl in Thorium-series with $Q_\beta$ is 5.001MeV. In the case of $^{210}$Tl, this is originated from α-decay of $^{214}$Bi with very low branching ratio (0.02%), so it is negligible background. Since these are beta-decays, energy of decay (Q-value) is shared randomly to beta particle and neutrino. This sharing energy process results in a continuous beta spectrum. In the case of $^{208}Tl$, its decayed beta particle forms a continuous spectrum ranging up to energy 5 MeV. Therefore, these decays can contribute as background in the Q-value region of $^{48}Ca$ (about 4.3 MeV).

Method for rejection: $^{208}Tl$ is originated from α-decay of $^{214}Bi$. Thus, to remove its β-decay, we can tag the preceding α-decay by knowing the half-life of $^{208}Tl$ (3 minutes). This method depends a lot on the dead-time of DAQ system. If the dead-time is smaller, tagging efficiency is higher. Beside minimized dead-time, position resolution is also important to reject this background.

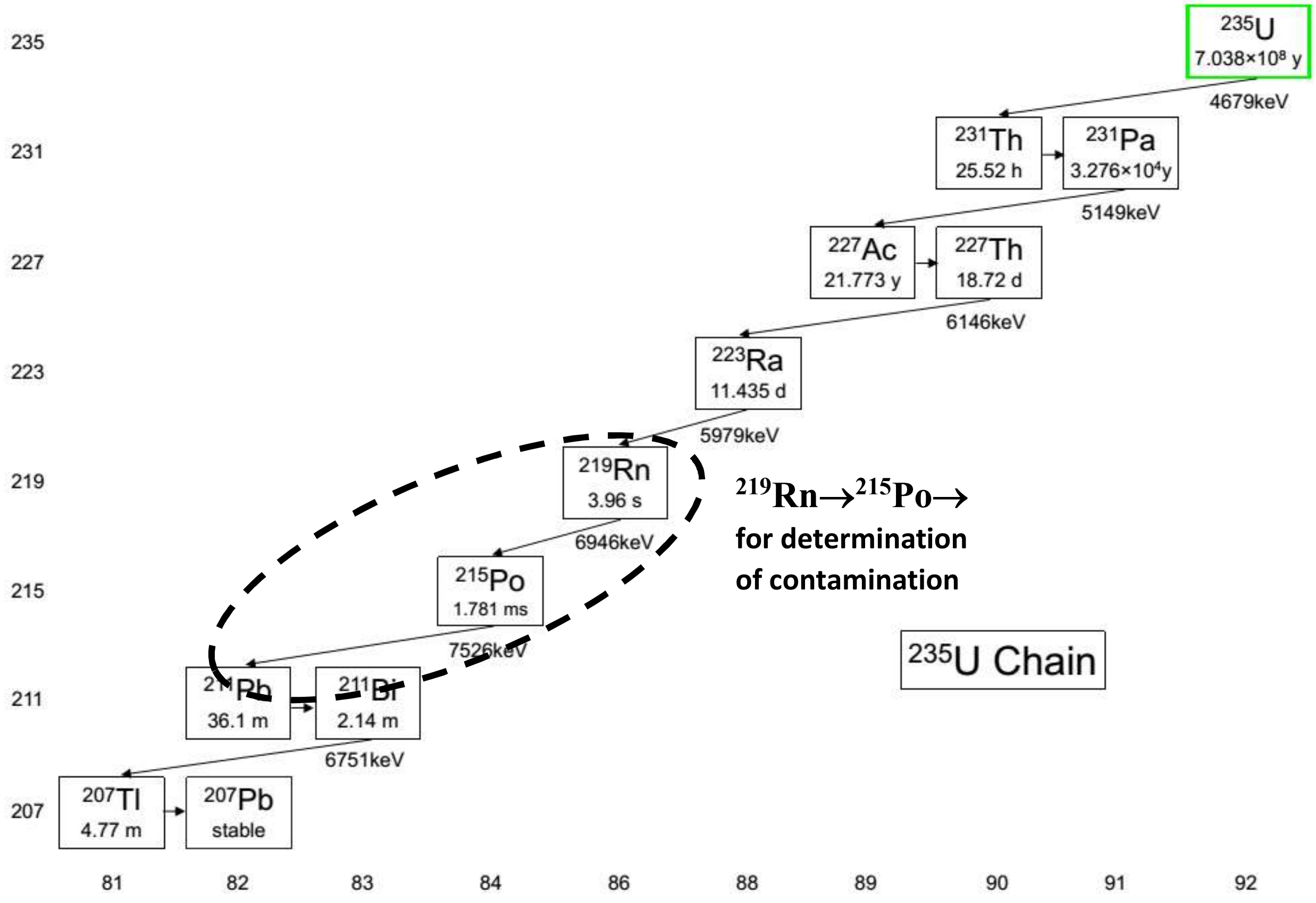

**Figure 2.8c.** Current decays used for determination of contamination ($^{219}Rn \rightarrow ^{215}Po \rightarrow$) originated from Ac-series are marked with black circle.

### c. Determination of contamination

Although we have methods for suppress the impurity backgrounds in U-series and Th-series, we need to confirm result with contamination measurement. Contamination can be determined from radioactivity of selected sequential decays which can be assumed secular equilibrium. Decays from impurities in Ac-series can effect determination of contamination analysis [31]. Thus, we also need to determine contamination of Ac-series. These decays are selected according to half-lives of daughter nuclei (less than 1 sec). Currently selected decays for determination of contamination are listed as following (and marked with black circle in Figure 2.8):

$$\text{U-series: } \underset{19.9\text{m}}{^{214}\text{Bi}} \xrightarrow{\beta} \underset{164\mu\text{s}}{^{214}\text{Po}} \xrightarrow{\alpha}$$

$$\text{Th-series: } \underset{55.6\text{s}}{^{220}\text{Rn}} \xrightarrow{\alpha} \underset{0.145\text{s}}{^{216}\text{Po}} \xrightarrow{\alpha}$$

$$\text{Ac-series: } \underset{3.96\text{s}}{^{219}\text{Rn}} \xrightarrow{\alpha} \underset{1.781\text{ms}}{^{215}\text{Po}} \xrightarrow{\alpha}$$

## 3. Requirement of DAQ system (Research Motivation)

Although Q-value of $^{48}$Ca is higher than most of natural backgrounds, there are several backgrounds from radioactive impurities whose decay energy is near interested energy, which is also Q-value of $^{48}$Ca – the energy region to distinguish 0νββ events from 2νββ events. Since these impurities lead to sequential decays, they have to be removed by tagging preceding and following events. One of significant backgrounds produced by impurities is beta-decays of $^{208}$Tl. Q-value of beta-decay from $^{208}$Tl (Q-value is about 5MeV), which is one of considered background in CANDLES. By tagging preceding alpha-decays originated from $^{212}$Bi and knowing half-life of $^{208}$Tl (3 min), we can remove these beta-decays. To observe 0νββ events, we have to reduce background as much as possible since event rate of 0νββ is extremely small. Therefore, tagging efficiency has to be high and, hence, dead-time of DAQ system should be minimized.

In 2016, a new DAQ system was introduced in CANDLES. It has a better performance of FADCs compared to the ones used in previous DAQ system. The FADC has 8 readout buffers and we use SpaceWire-to-GigabitEthernet network for readout data. Details of DAQ system is introduced in Chapter 3. My work is to realize the new DAQ system with minimized dead-time at 20 cps, which is trigger rate of CANDLES [32]. Goal for the work is to achieve DAQ efficiency in data taking, at least, equal to previous DAQ system. With efficiency about 98-99%, rejection of $^{208}Tl$ with previous DAQ system is about 60%. Additionally, the DAQ must obtain full waveform data, whose size is about 40 kB (512 Bytes × 74 FADC channels). With improved performance, it is expected to reduce more background within region of interested energy of CANDLES (Q-value 4.27 MeV).

# Chapter 3. DAQ system in CANDLES

The new DAQ system for CANDLES is described in this chapter. Hardware, firmware and software are newly developed. Content of this chapter is divided in two sections: hardware configuration and software configuration. Parallel reading is also explained in this chapter.

## 1. Hardware configuration

### 1.1. Micro-TCA hardware modules

#### a. Micro-TCA standard [33,34]

Micro-TCA (Micro Telecommunications Computing Architecture) is an embedded, scalable architecture with flexibility. Micro-TCA is chosen to build new DAQ system in CANDLES due to following reasons:

- Micro-TCA has Shelf Management function, which can control the power of the whole system. When overheating happens, it can be switched off automatically. This is helpful for safety in CANDLES since it is constructed underground.
- It uses point-to-point serial data link. This provides a high speed and flexible configuration.

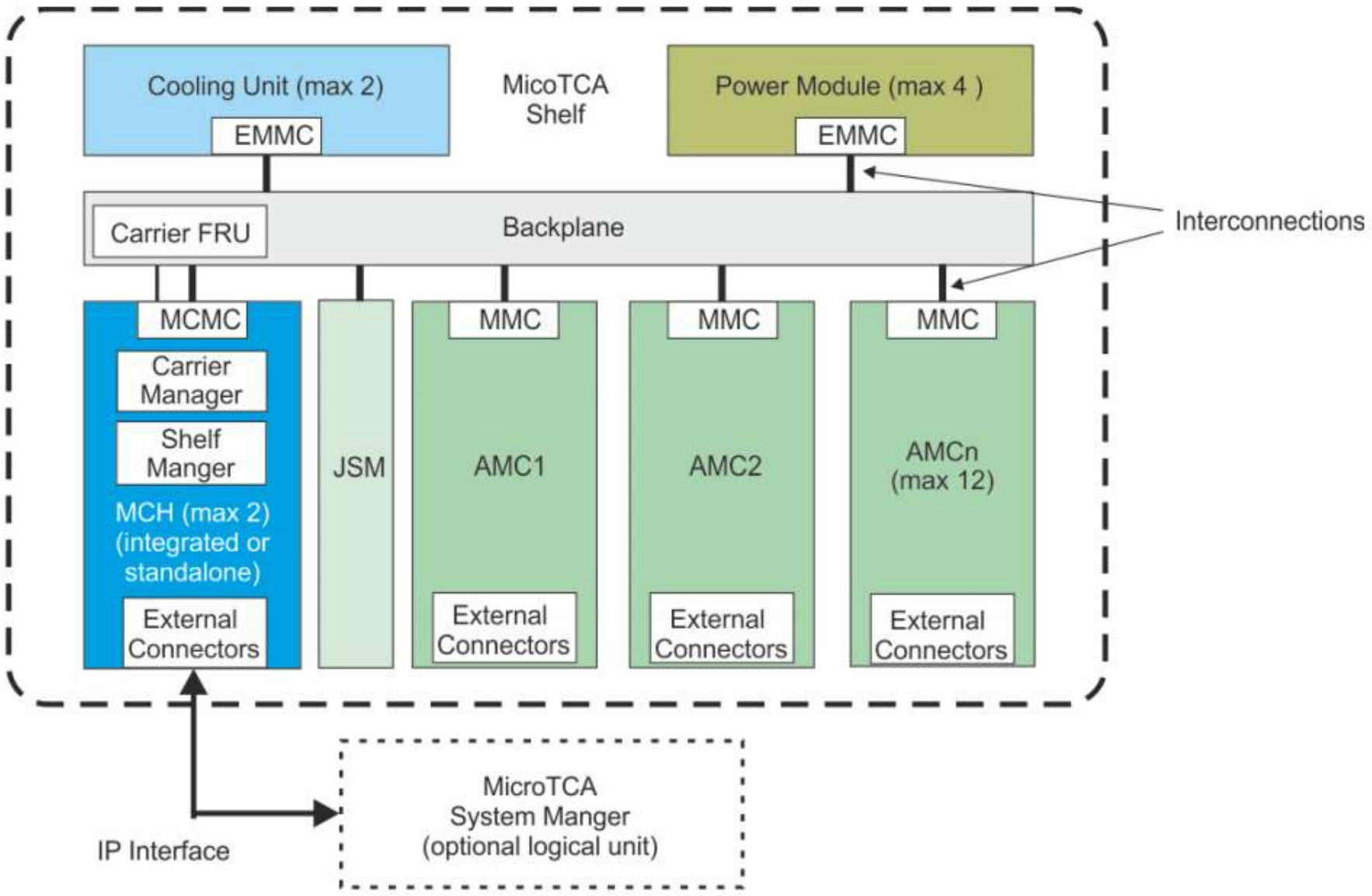

**Figure 3.1.** Block diagram of one typical Micro-TCA system [31]

A Micro-TCA system consists of crate, or chassis, Micro-TCA Carrier Hub (MCH), Advanced Mezzanine Card (AMC), Backplane, Cooling Units, Power Modules. (Figure 3.1). Brief introduction of these components are described as following:

• Micro-TCA crate: provides physical installation for other components. It also supports necessary management platform and sufficient air flow in Micro-TCA system.

• MCH: is the main management module. It provides required data connection and manages all AMCs, Cooling Unit, Power Modules in a Micro-TCA system. MCH plays an important role in data connectivity by switched fabric and clock distribution to all AMCs.

• Backplane: is a essential component in Micro-TCA system. Backplane allows point-to-point connections between MCH and AMCs, between AMCs. Up to 12 AMCs are connected to MCH.

• AMC: is the core component of Micro-TCA. In other words, they are the main reason for development Micro-TCA architecture. In Micro-TCA system, AMCs connect to backplane and they act as individual blades. AMCs are designed in order to support high speed connection.

**b. Micro-TCA hardware modules in CANDLES**

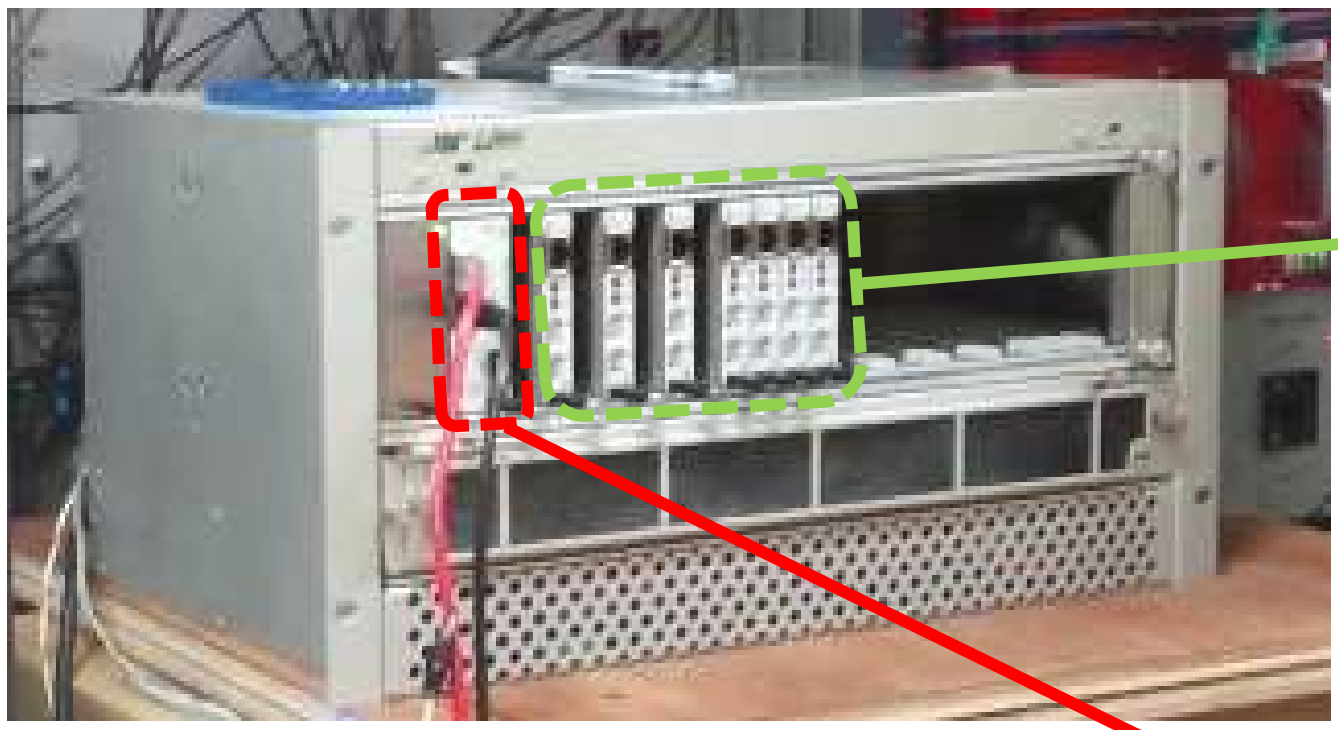

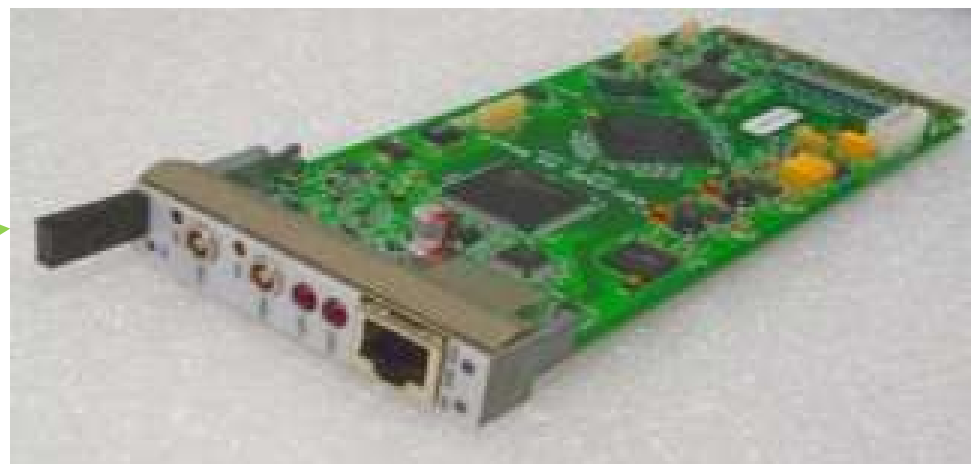

(b) AMC-FADC

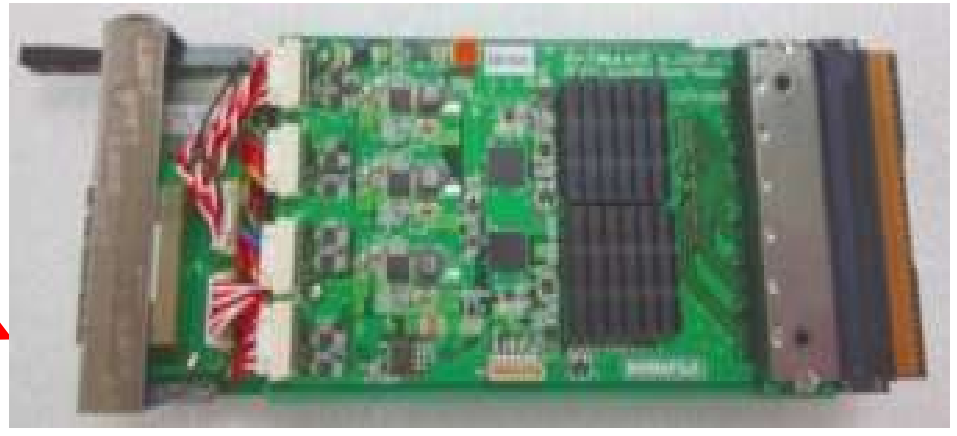

(a) Micro-TCA crate manufactured by Uber Ltd. including Power Modules, Cooling Units, Power Supply

(c) Micro-TCA Carriage Hub

**Figure 3.2.** Micro-TCA system used as DAQ in CANDLES experiment. Micro-TCA crate (a) contains AMC-FADCs (b) and Micro-TCA Carriage Hub (c)

**Table 3.1.** Details of Micro-TCA system used in CANDLES (extracted from [35])

| Module | Product of | FPGA logic | Details of FPGA development |
|---|---|---|---|
| MicroTCA crate | Uber | | |
| MCH | Shimafuji | Shelf Management and GbE-to-SpW interface | by Shimafuji |
| | | Space Wire Router | Open IP (by Shimafuji) |
| | | Trigger Controller for CANDLES | by Osaka University |
| AMC-FADC | Shimafuji | FADC control | by RCNP,<br>Osaka University |
| | | SpaceWire | Open IP (by Shimafuji) |

The Micro-TCA system in CANDLES is developed and produced to adapt requirements of CANDLES. Figure 3.2 is one MTCA crate including 1 MCH module and FADC modules. Details of FPGA logic (firmware) development and manufacturer are described in Table 3.1.

❖ **AMC-FADC**

FADC modules are designed as AMC blades. Each module has two input channels. Each channel has 8 readout buffers to reduce dead-time of DAQ system (this is discussed in section 1.2). Sampling rate of FADC channel is 500 MegaSample/sec (corresponding to 2 nsec/time-bin) with 8 bit resolution. The data is readout through SpaceWire network on the backplane. FADC has adjustable dynamic range with gain varies from 0.5 mV/channel to 4.5 mV/channel. SpaceWire is used not only for waveform readout but also for control and adjustment. By assignment value into designated registers, one can change trigger condition, adjust pedestal or timing, check available data, etc. AMC-FADC modules can be accessed easily from PC via MCH. Table 3.2 summarizes the specifications of AMC-FADC developed for CANDLES.

❖ **MCH**

Micro-processor chip controls power and fan of all elements in one Micro-TCA crate. MCH produced by Shimafuji consists of 3 FPGA chips (Figure 3.3):

- Clock distribution and GbE-SpW interface FPGA: It receives Global Clock from Master Module (which is described in following section) via RJ-45 port and distribute this clock to all AMC-FADCs via backplane. It provides GbE-SpW interface allowing Ethernet access from PC to modules of Micro-TCA system.
- SpaceWire Router FPGA: allows access from MCH all AMC-FADCs.
- Trigger FPGA: collects local trigger signals and busy signals from all AMC-FADCs. It distributes Global Trigger, Reset signals.

**Table 3.2.** Specifications of AMC-FADC

| | |
|---|---|
| Resolution | 8 bits |
| Sampling Rate | 500 MHz (2 nsec/time-bin) |
| Gain | 0.5~4.5 mV/channel (adjustable) |
| Offset | 0~255 (adjustable) |
| Event buffers (Readout buffers) | 8 buffers |

❖ **Number of crates, MCHs and AMC-FADCs**

As mentioned in Chapter 2, CANDLES consists of 62 PMTs for collecting scintillation photons. Signals observed by these PMTs are fed into FADCs for further digital pulse processing: trigger controlling (local and global), digitization, storing in event buffer and to be readout by PC. To observe all PMT signals, we use 37 AMC-FADCs in 4 Micro-TCA crates. Each crate containing 1 MCH. There are 74 FADC channels (2 channels/AMC). We use 62 channels for 62 PMTs are available, and the remaining 12 channels are used for trigger signals. Details of PMT-to-FADC mapping are indicated in Appendix A. All 4 MCHs provides GbE-SpW interface, PC on the Ethernet network can access and readout data from all FADCs.

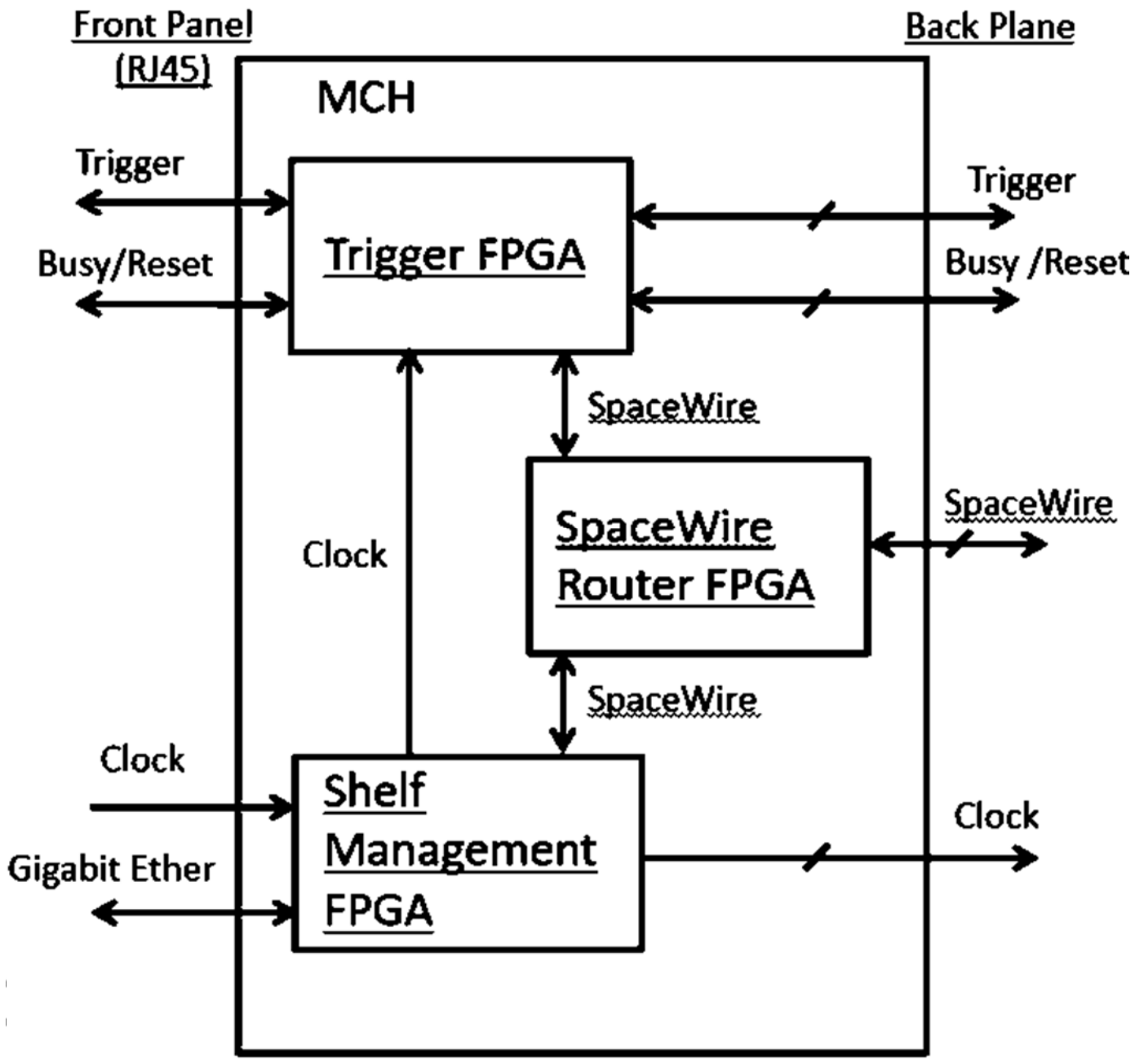

**Figure 3.3.** Block diagram of FPGA components inside MCH [36]

❖ **Master Module (for synchronization)**

Synchronization of FADCs is needed in further analysis of CANDLES. In our DAQ system, we use a Master Module to distribute global clock signals and global trigger signals.

• Global clock: Master Module distributes global clock signals to all 4 MCHs. After receiving global clock, each MCH distributes clock signals to all AMC-FADCs.

• Global trigger: The DAQ system is designed to readout data of all AMC-FADCs whenever trigger condition in each module is met. Trigger conditions set in each AMC-FADCs. A trigger signal is created when one of these conditions is met, and it is named Local Trigger. Each MCH collects Local Triggers from AMC-FADCs in the crate and dose the trigger decision. The result is transferred to Master Module. Master Module forms the Global Trigger from these trigger signals, and, then, distributes it to all AMC-FADCs via 4 MCHs.

• In addition to local trigger signals, busy signals from all modules are gathered by Master Module. Global is not distributed while gathered busy signals exist. The gathered busy signal indicates busy time of whole DAQ system. This helps to realize read-time measurement, which will be explained in later.

## 1.2. Trigger controlling and Event buffer

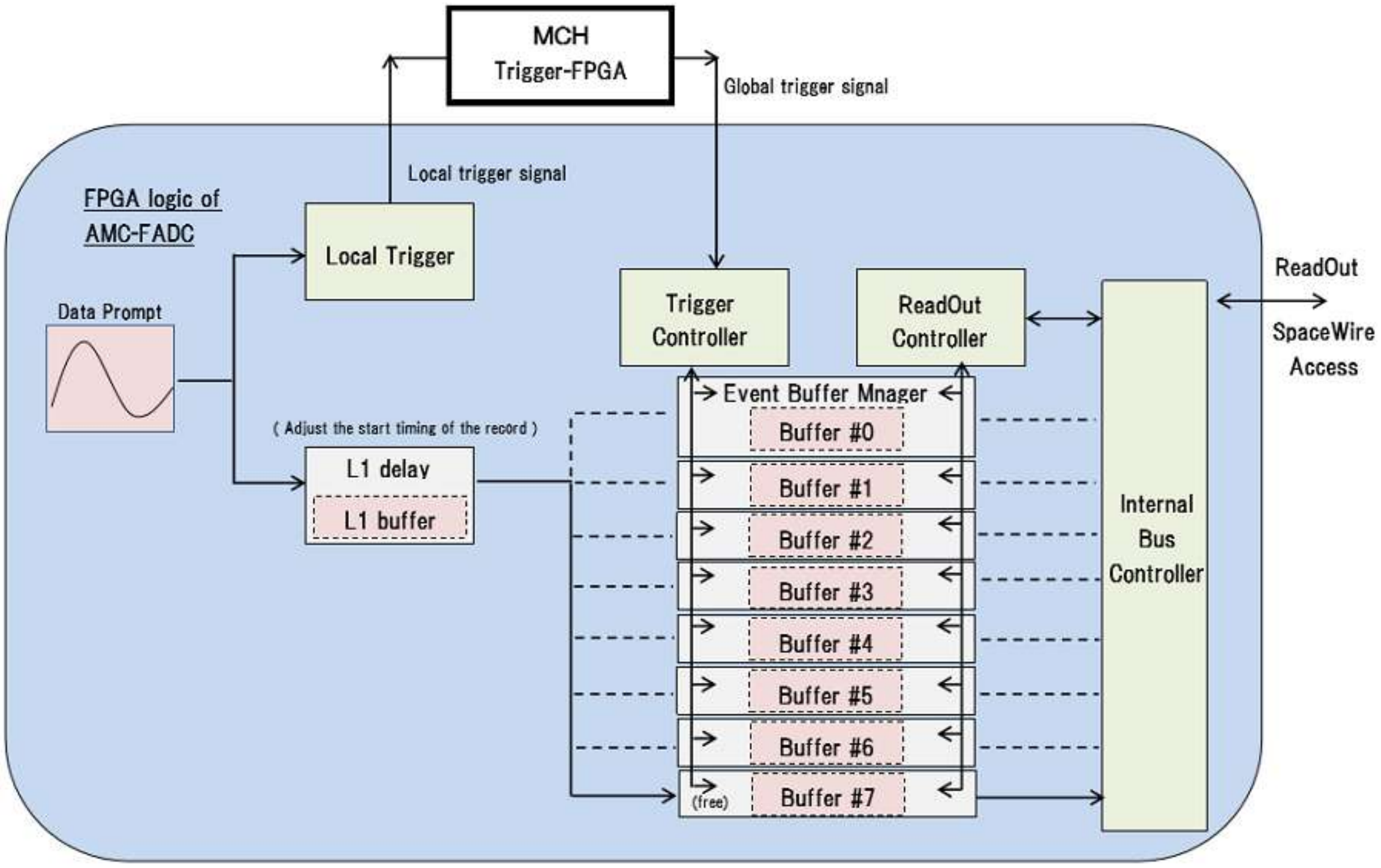

**Figure 3.4.** Schematic of trigger flow in DAQ system [36]

This section explains how waveform data is stored in Event buffer when AMC-FADC receives trigger signal. As abovementioned, there are two types of trigger in CANDLES: Local Trigger (produced by AMC-FADC) and Global Trigger (produced by Master Module). Local Trigger signal is set based on primitive analysis of the waveform at each FADC. Local Trigger signal is sent to MCH module. MCH gathers all Local Trigger signals in one crate. MCH does trigger decision using Local Trigger signals then send the result to Master Module. Trigger conditions consist of dual gate trigger, minimum bias, cosmic-ray, etc. Among these trigger condition, dual gate trigger is the primary trigger which selects $CaF_2$ signals [37]. The result of those trigger decision collected at the Master Module are distributed as a Global Trigger signal. MCH distributes the Global Trigger signals to FADC modules. On receiving global trigger, the FADC start to record waveforms into Event Buffer (8 buffers in total). By the time receiving global trigger, PC can access, check which buffer is ready and readout data from appropriate buffer. Event

Buffers can reduce DAQ inefficiency. The role of event buffer influence on DAQ efficiency is explained in section 2.c. of this chapter.

### 1.3. SpaceWire network

#### a. SpaceWire standard [38]

SpaceWire is a data link network designed to connect sub-systems onboard spacecraft. It provides high-speed (100Mbps of speed is used in CANDLES), bi-directional, full-duplex data. Network can be built to suit particular applications using point-to-point data links and routing switches. SpaceWire help us to build a flexible network for accessing and reading out data in SpW modules from PCs. SpaceWire protocol is simple, hence, it is easy to be implanted in FPGA.

❖ **RMAP**

RMAP is an abbreviation of Remote Memory Access Protocol. It was designed to support a wide range of SpaceWire applications. With SpaceWire network, we can access or connect devices using point-to-point link. RMAP is a protocol for accessing, reading, writing to registers or memory of remote devices within a SpaceWire network. Together, RMAP and SpaceWire provide a powerful data handling for various SpaceWire instruments.

#### b. SpaceWire network in new DAQ system

In CANDLES, SpaceWire is installed in FPGAs contained in AMC-FADCs, MCHs and Master Modules. All of these devices contain memories and registers which can be read or written by using RMAP. In order to access from PCs, which are used for data storage or online monitor. We developed a PCIe card allowing communication from PC to SpaceWire network. For convenience, in new DAQ system, we use Gigabit Ethernet to SpaceWire (GbE-SpW) interface, which is installed in MCHs. There are 4 TCP/IP ports in one MCH. Figure 3.6 is schematic diagram of data flow via SpaceWire network in new DAQ system. PC easily accesses to all 4 MCHs via Gigabit Ethernet, and then readout data in AMC-FADC event buffers via GbE-SpW interface and SpaceWire router. Gigabit Ethernet hub is used to access multi target from one PC. PC can write (for controlling) or read data from Master Module by using GbE-SpW converter (a product of Shimafuji [39]).

With GbE-SpW interface, it is possible to realize a DAQ system using SpacWire network with even an "off-the-shelf" computer. Additionally, Gigabit Ethernet provides a high speed network (1Gbps) higher than speed of SpaceWire network (100Mbps).

**c. Why choosing SpaceWire for CANDLES?**

There are two main reasons that we consider to apply SpaceWire in our DAQ system:

- At first, SpaceWire provides a flexible network. In several experiments where multiple detectors are used, each PC readouts data from one detector, then, data from these PCs are gathered by an Event Builder. Due to flexibility of SpaceWire, one PC in our DAQ system can readout data from all detectors. Thus, we do not need any Event Builder in our DAQ system. One example of other DAQ system with Event Builder is attached in Appendix B.

- Second reason is low cost development of SpaceWire standard. It is possible to build a network with only Gigabit Ethernet interface for easy access from PC to AMC-FADCs. However, we need large FPGAs for this purpose, thus, the cost for development is much higher compared to FPGAs for SpaceWire development.

## 2. Software configuration

### 2.1. SpaceWire/RMAP library

**a. Introduction**

Because we use GbE-SpW interface in our DAQ system, a SpaceWire/RMAP packet are embedded in TCP/IP packet. Readout PC needs to send or receive SpaceWire/RMAP packet which is blocked by header and footer of TCP/IP. Therefore, to realize the communication between PC(s) and SpaceWire network (i.e. DAQ system of CANDLES), we need a software which can:

- to construct SpaceWire/RMAP packets: in a convenient way to users
- to interpret received SpaceWire/RMAP packets: base on defined standard of RMAP to get data content (value of register, waveform data, etc.)
- to send and receive these packets through TCP/IP

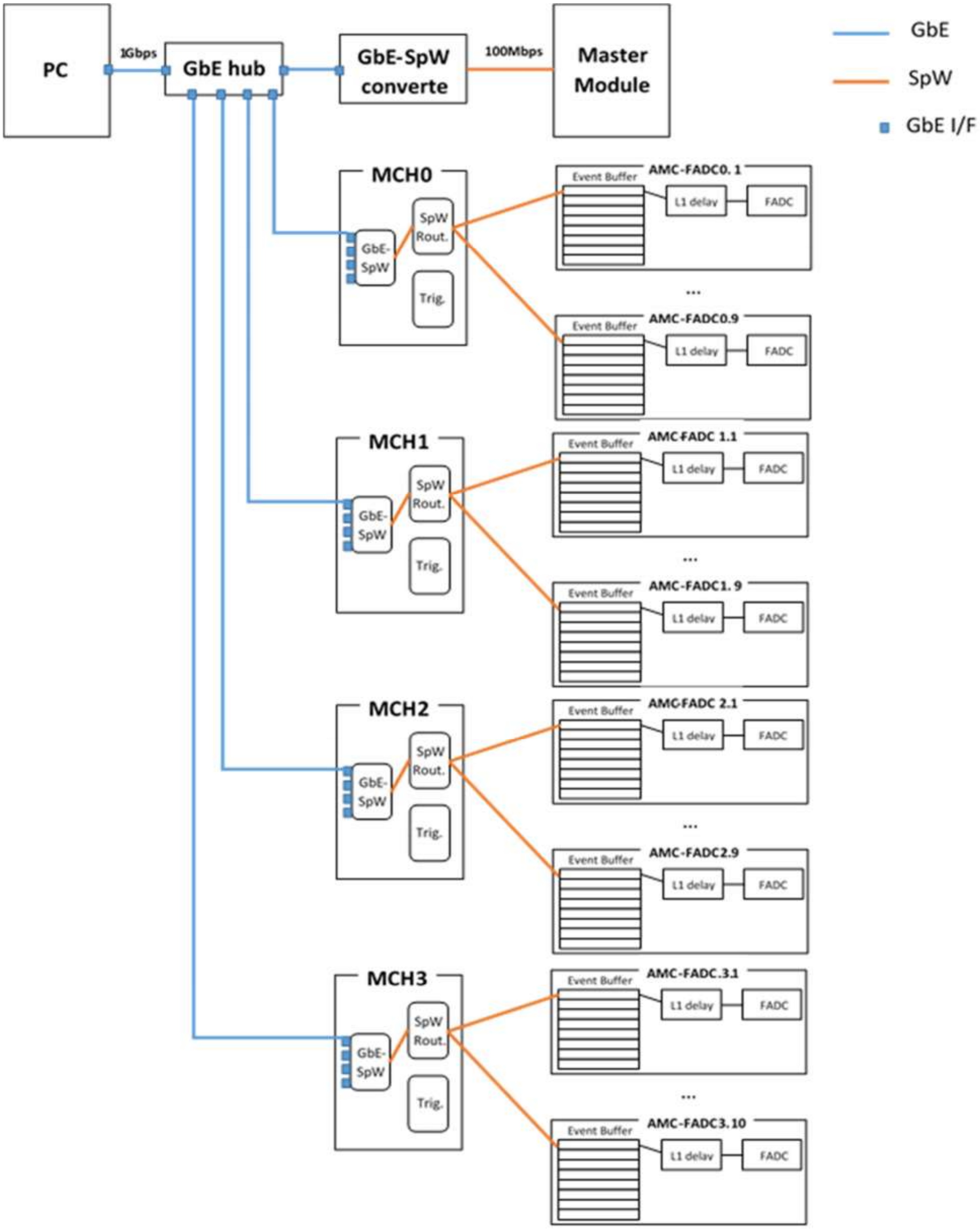

**Figure 3.5.** SpaceWire network in new DAQ system of CANDLES. SpaceWire network (100 Mbps) allows us to access AMC-FADCs from MCHs. GbE-SpW (1 Gbps) interface provides easy access to SpaceWire network with any “off-the-shelf” PC.

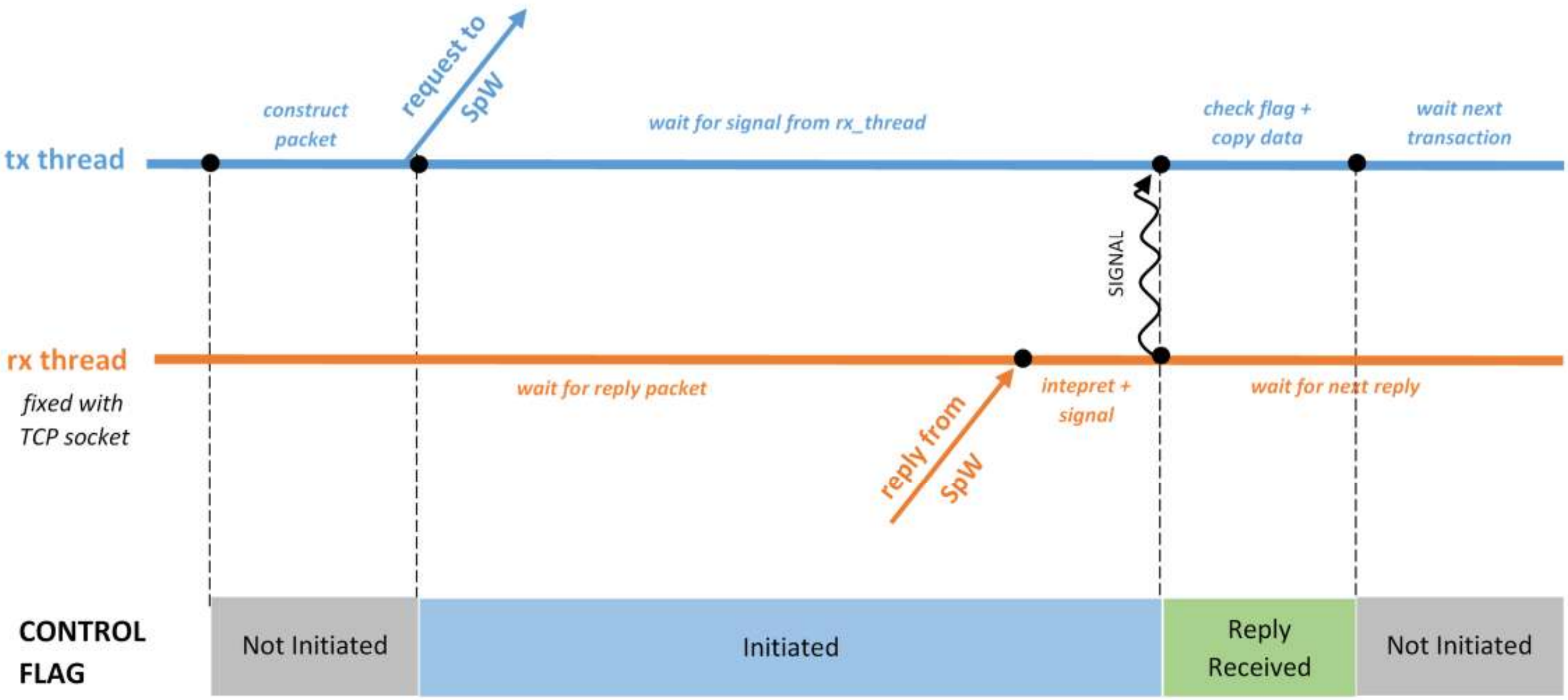

**Figure 3.6.** Mechanism for read/write data with SpaceWire/RMAP library

"SpaceWire RMAP library" [40] written by Takayuki Yuasa (JAXA) was chosen for this purpose. It is an open source C++ software and highly modularized. Besides requirements mentioned above, it also provides XML-like configurations allowing easy handling huge SpaceWire network with numerous target nodes.

### b. RMAP transaction for read/write data

#### ❖ Introduction

Communication in SpaceWire standard is based on sending request and receiving reply packet. Reply packet of reading command contains data, while reply packet of writing command confirms command executed. In SpaceWire/RMAP library, sending process and receiving process are distributed in two independent threads, which are simultaneous processes in PC programming. In this thesis, threads of sending process and receiving are named tx_thread and rx_thread, respectively. These two threads are plotted in Figure 3.6: tx_thread is plotted in blue and rx_thread is plotted in orange. Because a SpaceWire/RMAP packet is encapsulated in TCP/IP packet. TCP socket is used, and this socket is fixed with rx_thread. This means rx_thread receives all packets transferred in this socket. A set of "control flags" are used to inform states of transaction and build a "handshake" between these two threads.

❖ **RMAP transaction**

RMAP read transaction is explained below. At the beginning, tx_thread and rx_thread are set, "control flag" as "Not Initiated" (transaction is not created). When read/write command is executed, tx_thread first constructs SpaceWire/RMAP packet. The packet is sent via TCP socket to SpW-GbE interface modules. After sending, tx_thread "falls to sleep" and wait for a wake-up signal from rx_thread. "Control flag" now is changed to "Initiated". Rx_thread keeps waiting for reply packets in TCP socket.

Data reached PC is interpreted to get information of interest. When this process is finished, rx_thread creates a "SIGNAL" to wake up the tx_thread and, then, waits for next reply packet. "Control Flag" now is shifted to "Reply Received" state. After a wake-up-call, tx_thread first confirms whether the current state is "Reply Received". When transaction state is verified, information of interest is copied for further process. This is the end of transaction. Tx_thread is then waits for next reading command to execute, and transaction state is reset to first state, which is "Not Initiated". These processes and transaction states are repeated during data taking.

**c. Number of TCP sockets and threads for new DAQ system**

As discussion, TCP sockets are the key to communicate between PC and SpaceWire modules in DAQ system because data are transferred/received via these sockets. We can see in Figure 3.5 that it needs 5 sockets to connect Master Module and 4 MCHs. Every TCP socket is mounted with one rx_thread and, at least, one tx_thread.

*First byte transmitted*

| | | | |
|---|---|---|---|
| | Source Path Address | Source Path Address | Source Path Address |
| Source Logical Address | Protocol Identifier | Packet Type, Command, Source Path Addr Len | Status |
| Destination Logical Address | Transaction Identifier (MS) | Transaction Identifier (LS) | Reserved = 0 |
| Data Length (MS) | Data Length | Data Length (LS) | Header CRC |
| Data | Data | Data | Data |
| Data | Data | Data | Data |
| Data | Data CRC | EOP | |

*Last byte transmitted*

**Figure 3.7.** Format of read reply packet [41]

## 2.2. DAQ Middleware framework

### a. Introduction

DAQ Middleware [39] is a framework developed in KEK (High Energy Accelerator Research Organization). DAQ Middleware has implemented DAQ functions as separated components. All DAQ components can be executed in a separated computer, hence, it is possible to build a distributed DAQ network containing different components. These components run as basic functions for DAQ, therefore, it is able to construct a DAQ system using DAQ Middleware.

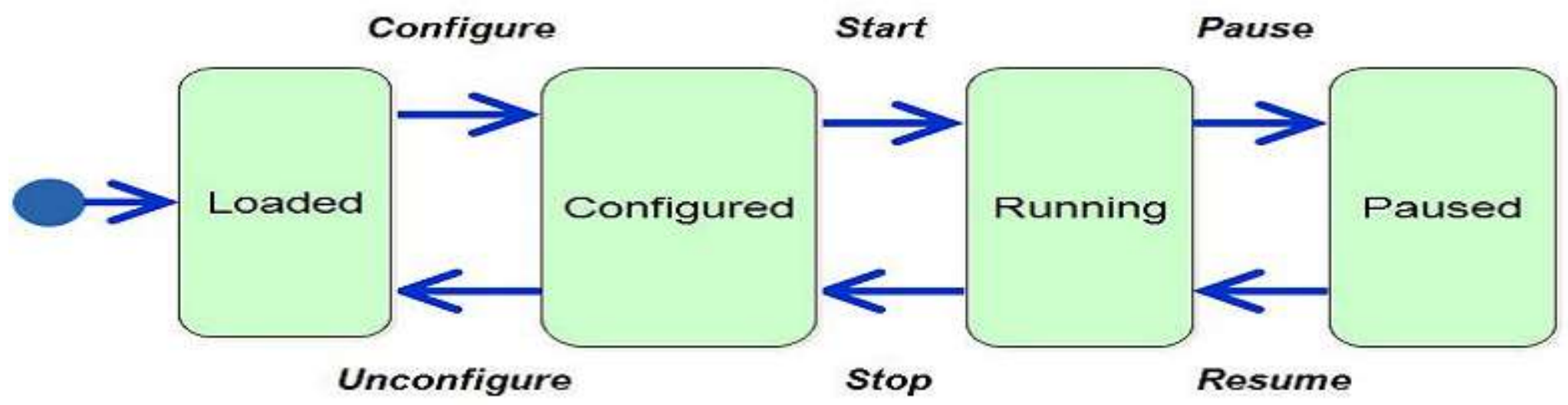

**Figure 3.8.** Diagram of state transition of DAQ Middleware [42]

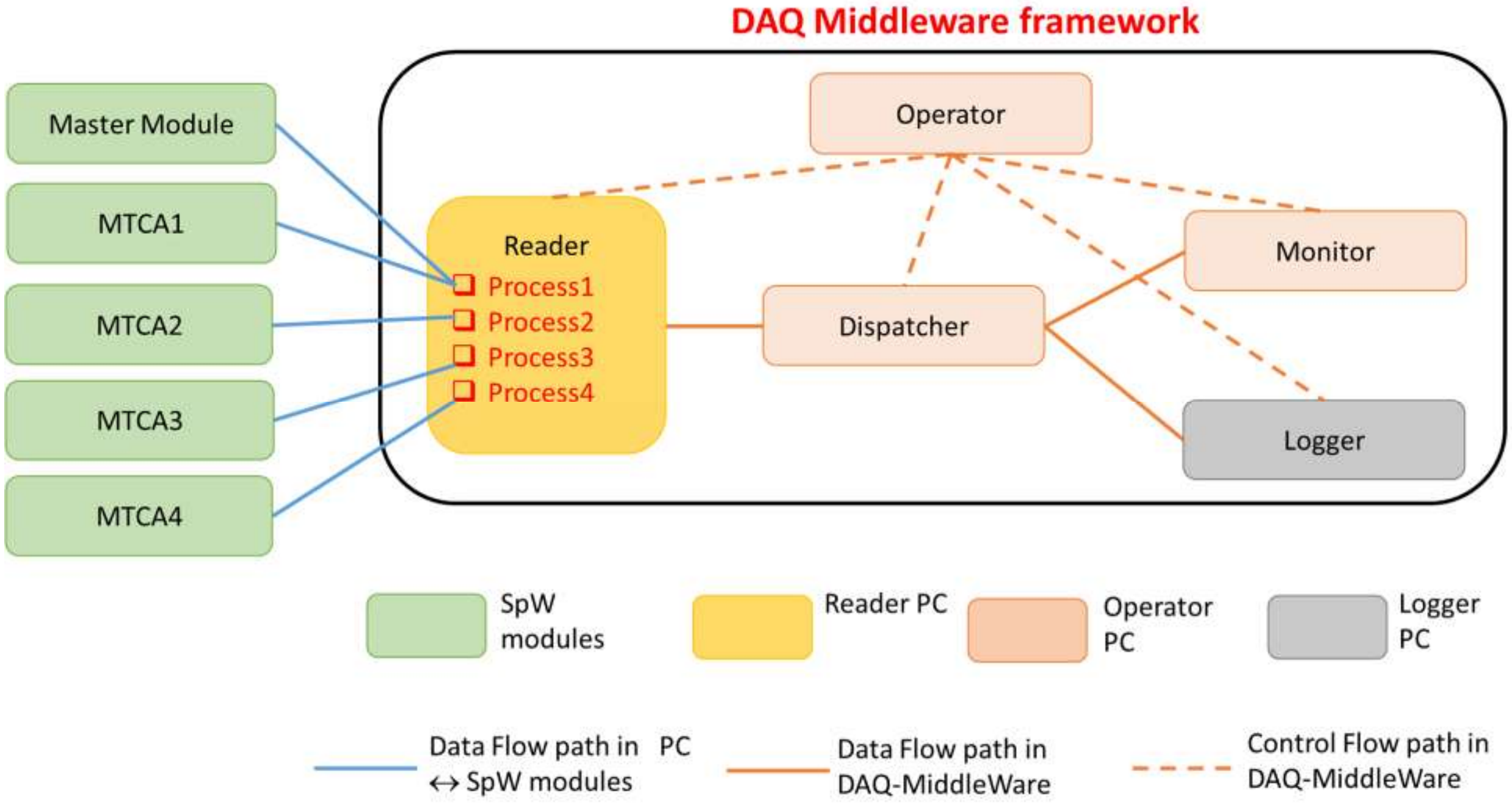

**Figure 3.9.** DAQ Middleware configuration in CANDLES. Three PCs are used to distribute DAQ network. All PCs and SpaceWire modules are connected via one GbE hub.

The DAQ Middleware consists of components: Operator, Dispatcher, Logger, Monitor and Reader. Every DAQ component connects to each other. In-Port of one component is called to receive data from Out-Port of another component. Additionally, each DAQ component offers state transition and activities (processes defined by user). States of every component in DAQ Middleware are expressed in Figure 3.8. Components of DAQ Middleware can be distributed in different machines (PCs). Reader component receives data stream observed from detectors, and forwards to Dispatcher. Dispatcher, then, transfers data stream to Logger Component and Monitor Component. Logger Component records data into hard disk of PC. Monitor Component displays data on histogram, which is updated continuously, using ROOT framework. Operator Component can control the whole DAQ Middleware system.

**b. DAQ Middleware in CANDLES**

All SpaceWire modules with GbE-SpW interface are connected to GbE hub (as shown in Figure 3.5). PCs in new DAQ system are connected to the same hub. Three PCs are used to distribute DAQ Middleware components:

- Reader PC contains Reader Component. It reads/writes data from/to SpaceWire modules of DAQ system.
- Logger PC contains Logger Component. It stores data stream to its own hard disk.
- Operator PC contains other components: Operator, Dispatcher and Monitor. It controls the whole DAQ system, forwards data stream to Logger PC and displays interested data dynamically with ROOT histogram.

This DAQ network is reused from previous DAQ system [32]. In previous DAQ system, we used Advanced-TCA based system with SpaceWire-to-PCIe interface, while the new DAQ system uses GbE-SpW interface. Because of the difference of the data readout interface, we have to modify the Reader Component to be applicable to Micro-TCA system. Reader Component reads/writes data from/to Master Module, 4 MCHs and 37 AMC-FADCs. In the Reader Component, we implement 4 processes associated with data transfer/receive of 4 MTCAs consisting of 4 MCHs and 37 FADCs. Since the data from Master Module is very small compared to data of 1 MTCA, process 1 is in charge of data communication of both MTCA1 and Master Module. Every other process is in charge for data from/to one MTCA. Details of DAQ Middleware configuration are indicated in Figure 3.9.

### 2.3. Parallel data readout

#### a. Data rate in GbE-SpW network

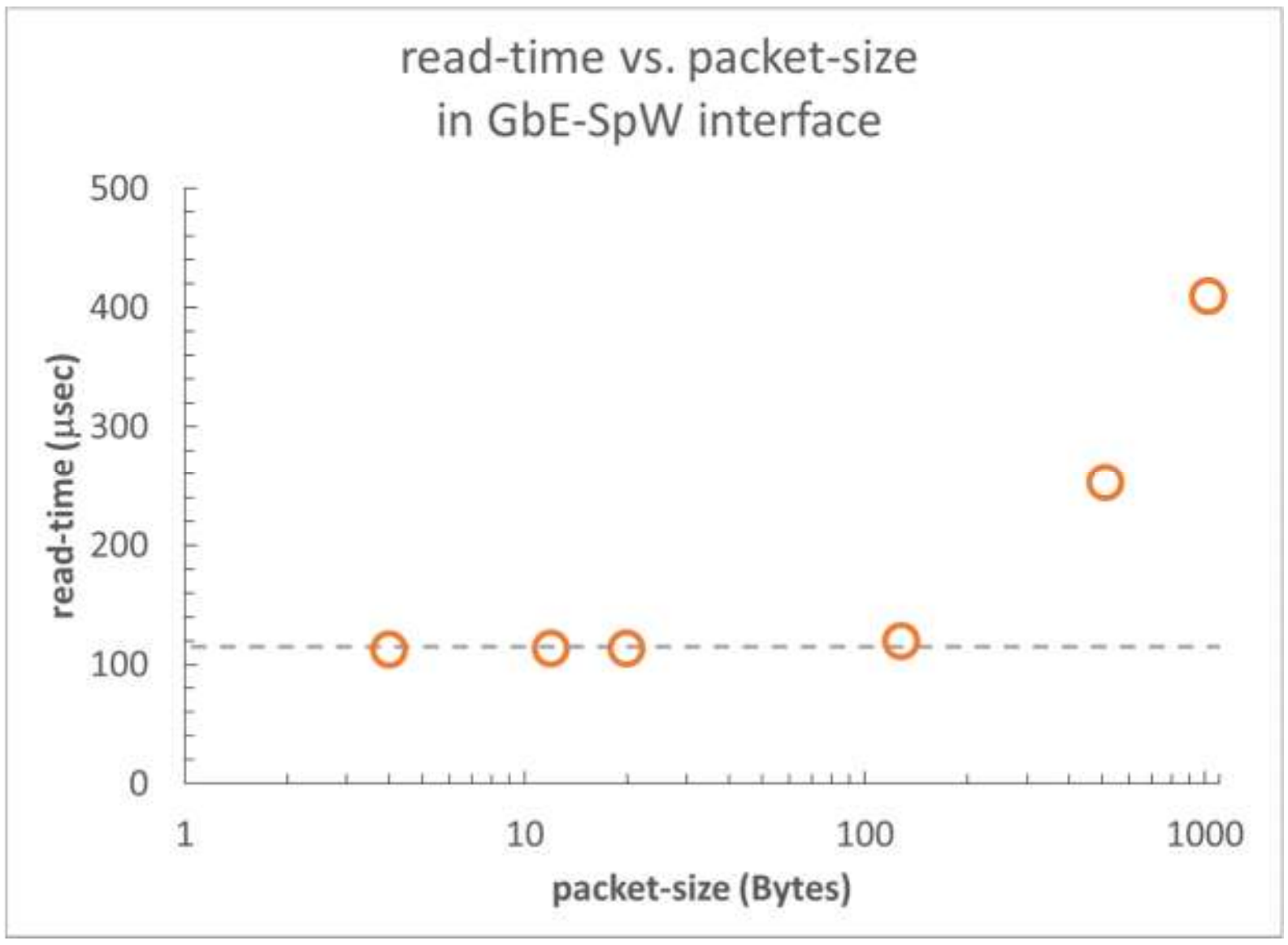

**Figure 3.10.** Measured read-time in GbE-SpW interface of various packet size. Dash line indicates constant overhead occurs with even small package size.

Figure 3.10 shows the read-time corresponding to several packet sizes. Size of waveform data obtained from FADC is 512 Bytes. Read-time at waveform data size is about 250 µsec. Read-time of header in FADC packet (12 Bytes) or any packet size below 100 Bytes is about 100 µsec. This is because large overhead caused by TCP/IP. Data transferred via GbE-SpW interface is sparse. Thus, in order to overcome the huge overhead of TCP/IP when using GbE-SpW interface, Reader component is developed with parallel data readout.

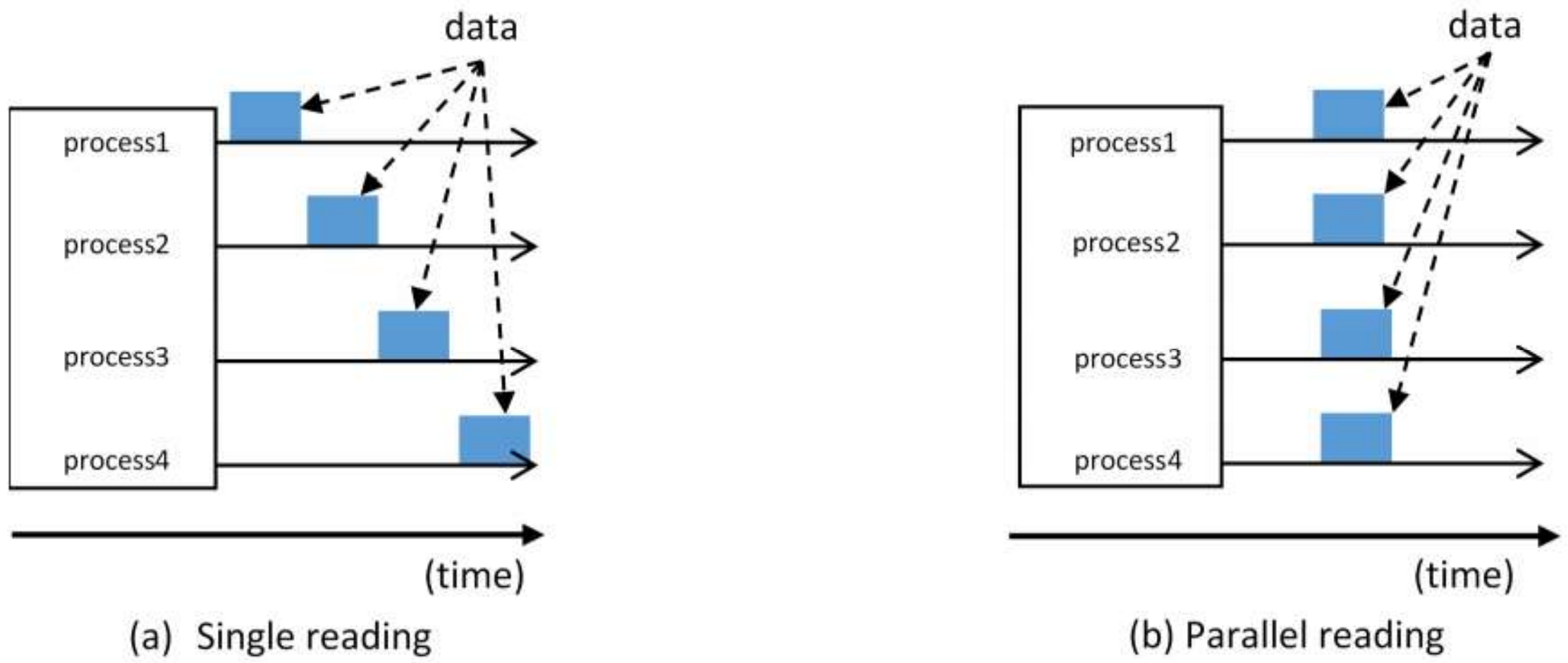

**Figure 3.11.** Flow of data stream of single reading (left-handed) and parallel reading with 4 parallel processes (right-handed). Data in single reading are readout sequentially, while data in parallel reading are readout simultaneously.

**b. Parallel readout in new DAQ system**

As mentioned in section 2.2, there are 4 processes implemented in Reader Component of DAQ Middleware, and these processes are in charge of data readout from Master Module and modules in 4 Micro-TCAs (including 4 MCHs and 37 AMC-FADCs). Moreover, Gigabit Ethernet interfaces of these SpaceWire modules are independent to each other. In order to construct a parallel readout, instead of executing these processes one by one, we modify the software in Reader Component to run these 4 processes at the same time by using multithread programming based on C++ language. After confirming data in AMC-FADCs are ready to read, Reader PC simultaneously access and readout data from 4 Micro-TCA crates. In addition, considering transferring speed, maximum speed of SpaceWire is 100 Mbps. With independent Gigabit Ethernet interfaces, even if there is conjunction in parallel readout, maximum speed of data transfer is about 400 Mbps. This maximum speed is lower than bandwidth allowance of Gigabit Ethernet (1Gbps). Figure 3.11 is an illustration of single reading and parallel reading processes using Reader PC in new DAQ system. In single reading, data stream from each Micro-TCA are readout sequentially. On the other hand, in parallel reading, data are readout simultaneously.

**c. Influence of read-time and multi event buffers on DAQ inefficiency**

In this section, we will see how multi event buffers works in DAQ system. If frequency of incident events is μ and read-time of DAQ system is T, average number of events coming within read-time is μT. Since the read-time T is in the order of milliseconds and trigger rate is about tens of events, average number of events within read-time T is small. Therefore, most of the time, buffers are empty. Inefficiency, which is named as $Q_N(\mu T)$ where N is number of event buffers, is the probability that more than N events occurred proceeding time T. Thus, $Q_N(\mu T)$ is defined as:

$$Q_N(\mu T) = 1 - \sum_{n=0}^{N-1} P(n, \mu T) \quad (3.1)$$

where P(n,μT) is the Poisson distribution to obtain n events with average μT events:

$$P(n, \mu T) = \frac{e^{-\mu T} (\mu T)^n}{n!} \quad (3.2)$$

Since μT is small, Taylor expansion of P(n,μT) is deduced:

$$\frac{d}{d(\mu T)} P(n, \mu T) = \frac{-e^{-\mu T} (\mu T)^n}{n!} + \frac{e^{-\mu T} (\mu T)^{n-1}}{(n-1)!} \quad (3.3)$$

$$\frac{d}{d(\mu T)} P(n, \mu T) = -P(n, \mu T) + P(n-1, \mu T) \quad (3.4)$$

In case of P(0,μT):

$$P(0, \mu T) = e^{-\mu T} \quad (3.5)$$

$$\frac{d}{d(\mu T)} P(0, \mu T) = -e^{-\mu T} = -P(0, \mu T) \quad (3.6)$$

Therefore, we have:

$$\frac{d}{d(\mu T)} Q_N(\mu T) = -\frac{d}{d(\mu T)} \left[ P(0, \mu T) + P(1, \mu T) + P(2, \mu T) ... \right] \quad (3.7)$$

$$\frac{d}{d(\mu T)} Q_N(\mu T) = -\left[ -P(0, \mu T) + P(0, \mu T) - P(1, \mu T) + P(1, \mu T) - P(2, \mu T) + ... \right] \quad (3.8)$$

$$\frac{d}{d(\mu T)} Q_N(\mu T) = P(N-1, \mu T) \quad (3.9)$$

We can derive higher power of derivative:

$$\frac{d^2}{d(\mu T)^2} Q_N(\mu T) = \frac{d}{d(\mu T)} P(N-1,\mu T) = P(N-2,\mu T) - P(N-1,\mu T) \quad (3.10)$$

$$\frac{d^3}{d(\mu T)^3} Q_N(\mu T) = P(N-3,\mu T) - 2P(N-2,\mu T) + P(N-1,\mu T) \quad (3.11)$$

$$\frac{d^4}{d(\mu T)^4} Q_N(\mu T) = P(N-4,\mu T) - 3P(N-3,\mu T) + 3P(N-2,\mu T) - P(N-1,\mu T) \quad (3.12)$$

Repeating these processes times by times, we have:

$$\frac{d^N}{d(\mu T)^N} Q_N(\mu T) = P(0,\mu T) - ... \quad (3.13)$$

We can easily see that $P(0,\mu T) = 1$ and $P(n>0,\mu T) = 0$. Therefore, Taylor expansion of $Q_N(\mu T)$ can be deduced:

$$Q_N(\mu T) = \frac{(\mu T)^N}{N!} \quad (3.14)$$

Equation (3.14) shows inefficiency of an N-buffers system as a function of $\mu T$. With same $\mu T$, inefficiency of an N+1 buffers system is reduced. We can compare the ratio of inefficiency between two systems:

$$\frac{Q_{N+1}(\mu T)}{Q_N(\mu T)} = \frac{(\mu T)^{N+1}}{(N+1)!} \Big/ \frac{(\mu T)^N}{N!} = \frac{\mu T}{N+1} \quad (3.15)$$

For instance, with $\mu T$ is 0.1, inefficiency of 4-buffers system is 1/40 (0.025) times smaller than 3-buffers system. To achieve the same order of reduction but not change number of event buffers, it is needed to reduce $\mu T$:

$$\frac{Q_3(\mu' T')}{Q_3(\mu T)} = \left(\frac{\mu' T'}{\mu T}\right)^3 = 0.025 \Rightarrow \frac{\mu' T'}{\mu T} = 0.025^{1/3} \approx 0.3 \quad (3.16)$$

In this example, to achieve the same inefficiency, it is needed to reduce $\mu T$ down to 3 times or just by adding 1 buffer. Figure 3.12 is estimations of inefficiency as a function of $\mu T$ are plotted with various number of buffers. In the new DAQ system, we combine the reduction of read-time and increasing number of buffers to minimize the inefficiency of DAQ system.

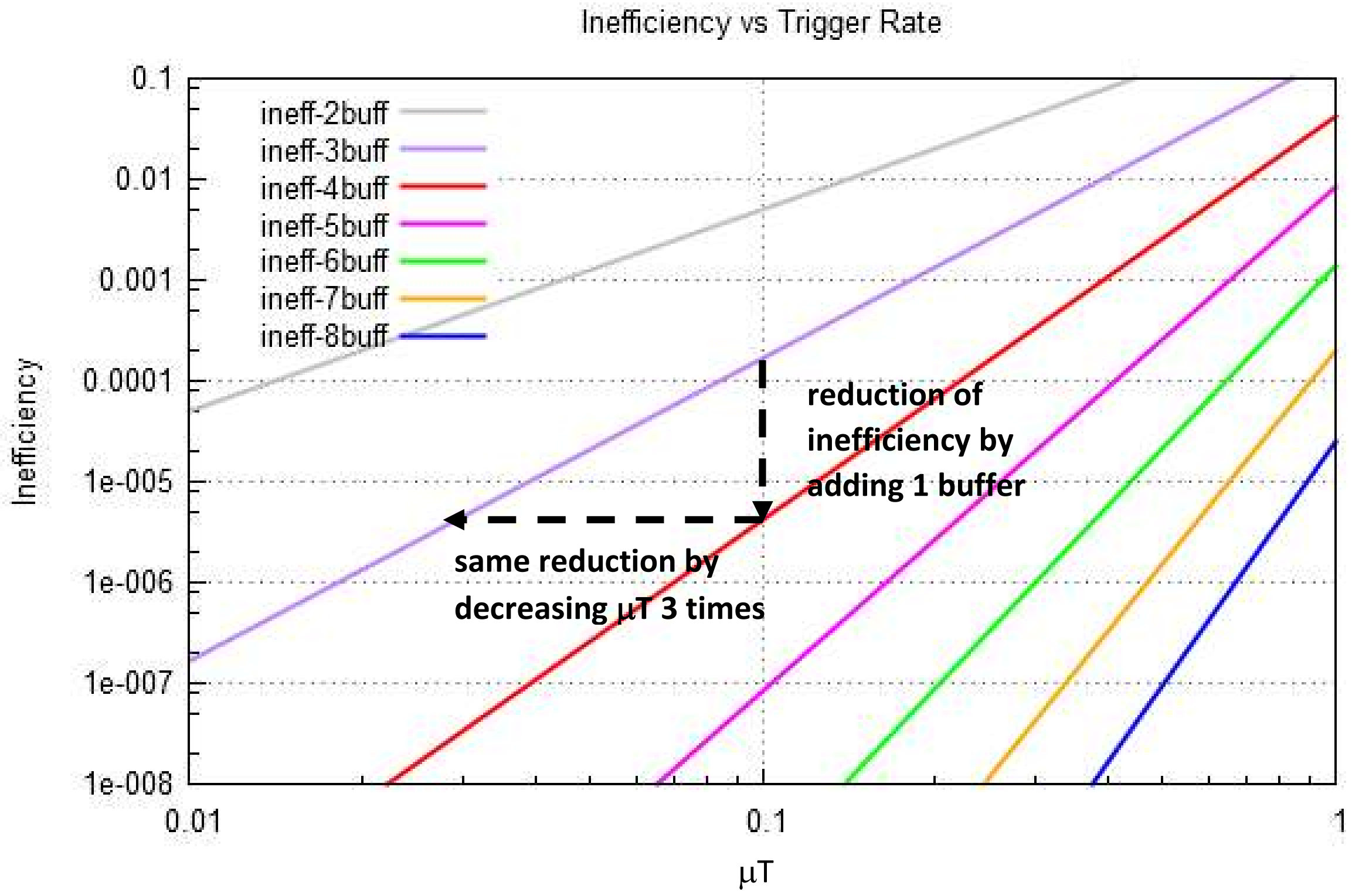

**Figure 3.12.** Inefficiency corresponding to trigger rate with various number of buffers

# Chapter 4. Evaluation of DAQ performance

In this chapter, measurements for evaluation of DAQ performance are described. The experiments include parallel reading test, inefficiency with multi event buffers and evaluation of DAQ efficiency of data taking. The final performance of new DAQ system is compared with previous DAQ system. Finally, tagging efficiency with new DAQ system is discussed.

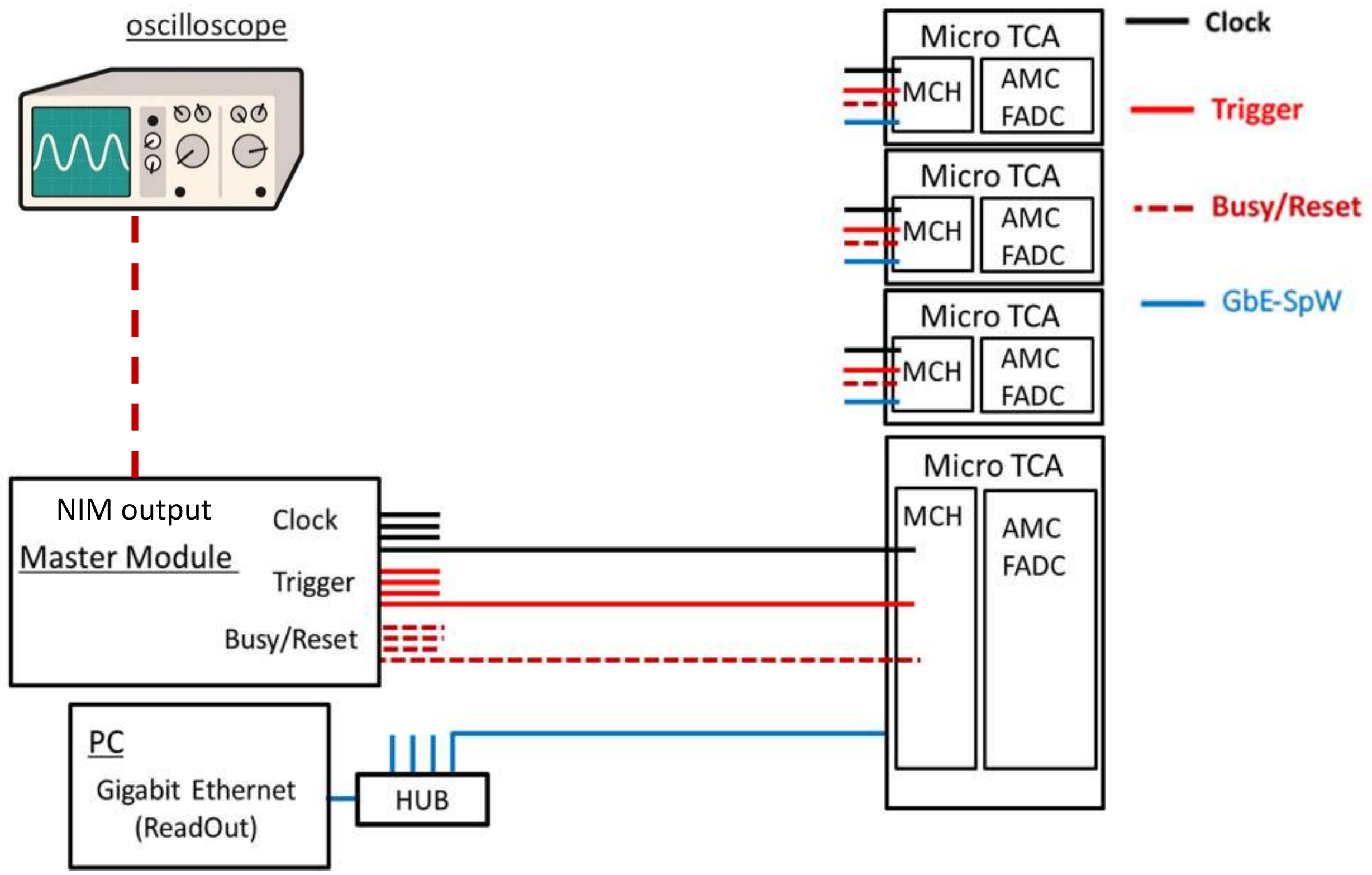

**Figure 4.1.** Experiment set up for measuring read-time

## 1. Parallel reading to reduce read-time

Test of parallel reading discussed in Chapter 3 is described in this section. Figure 4.1 is the set up for this test. As mentioned in previous chapter, Master Module gathers trigger and busy signals from all 4 MCHs. Thus, the width of gathered busy signal indicates the dead-time of the whole DAQ system. Since dead-time of other processes (gather triggers, gather busy, etc.) are negligible small, to reduce dead-time, we realize parallel readout. For evaluation the read-time of single reading process and parallel reading process, we measure the width of gathered busy signal. Additionally, this gathered busy signal can be taken out for measurement via NIM (Nuclear Instrument Modules) logic output of Master Module.

In this test, an input signal with 1cps of regular frequency generated from function generator is used. Software of Reader Component is modified in order to execute single reading (1 thread) and parallel reading (2 threads and 4 threads). Busy signal of AMC-FADC is generated only when all available buffer(s) are filled with event data. Number of event buffer in this test is set as one, and busy signal is set while readout process going. Busy signal of the whole DAQ system is observed and measured with an oscilloscope. A digital oscilloscope DSO7104B [43] (Agilent Technologies) with 4 GSa/sec of sampling rate and 1 GHz bandwidth is used for measuring. Width of busy signal indicating corresponding read-time are measured with this set up. Oscilloscope supports function to obtain mean, standard deviation, number of observed events. Read-time to obtain full event data size of CANDLES of 1 thread, 2 threads and 4 threads are measured. Figure 4.2 expresses captures of busy signals (of single reading and 4 parallel readings) on oscilloscope screen.

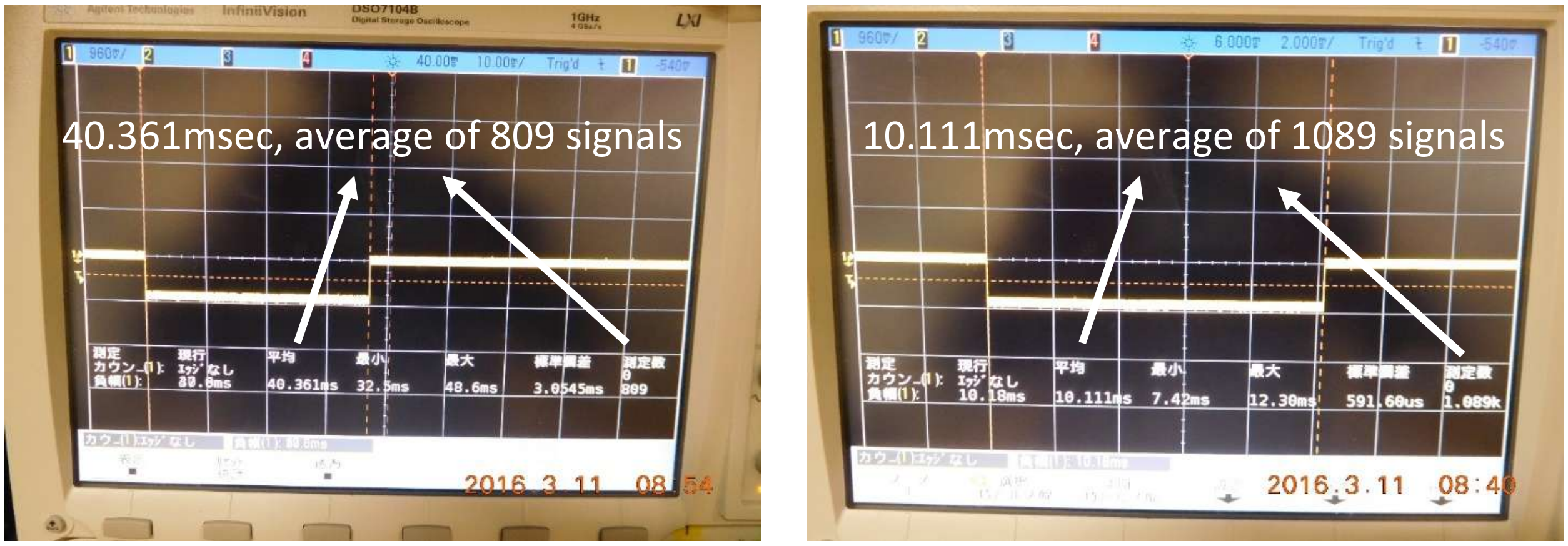

**Figure 4.2.** Observed busy signal of single reading measurement (left) and 4 parallel reading measurement (right). With 4 parallel reading, read-time is reduced 4 times (40.361msec down to 10.111msec). Obtained data includes mean, min, max, standard deviation and number of detected signals.

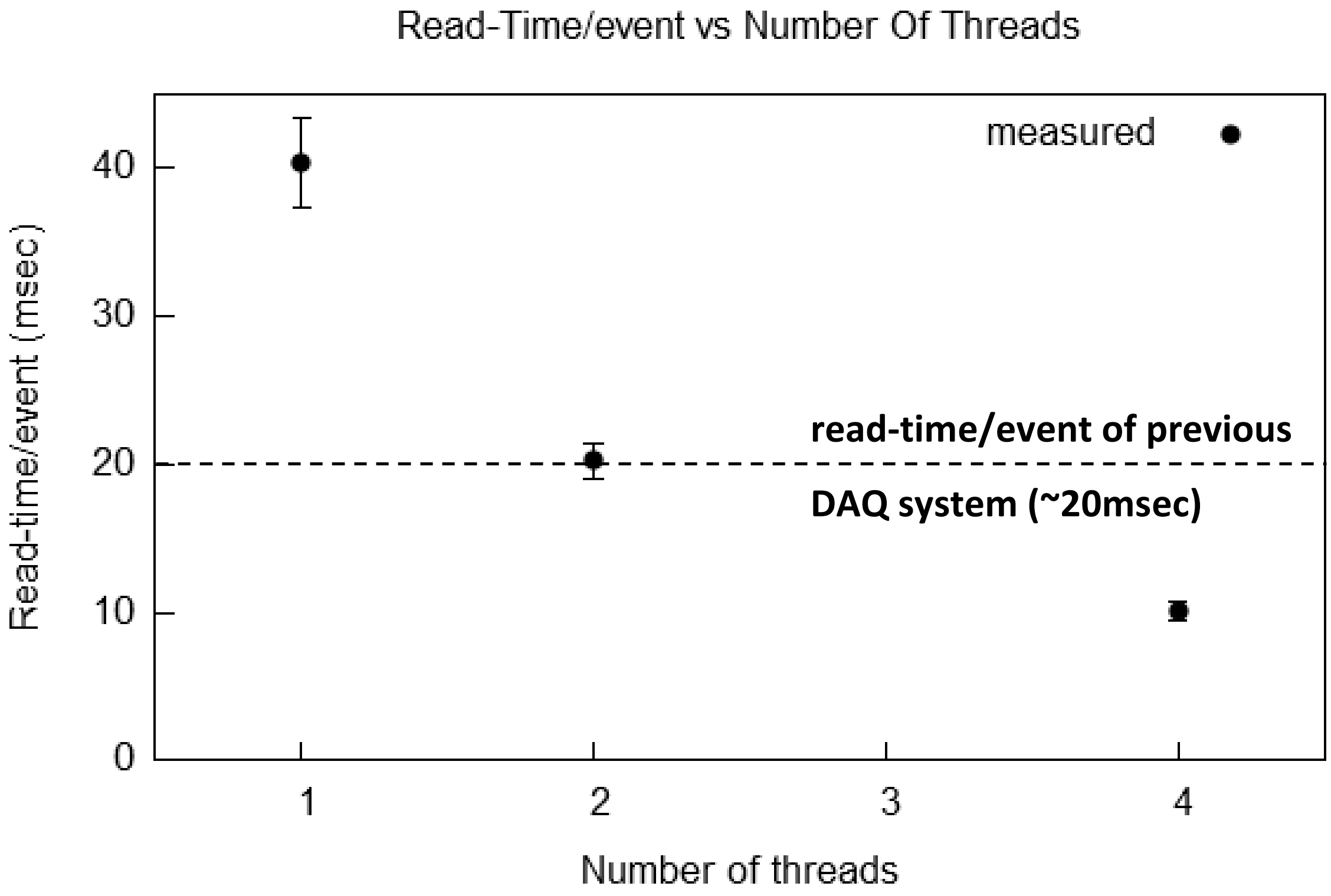

**Figure 4.3.** Obtained read-time for event data size of new DAQ system. Measured data are plotted with circles. Read-time/event of previous DAQ system is indicated with dash line.

**Table 4.1.** Read-time to obtain one event data size with new DAQ system

| Number of threads | Read-time/event (msec) | Standard deviation (msec) | Observed events |
|---|---|---|---|
| 1 | 40.361 | 3.054 | 809 |
| 2 | 20.236 | 1.214 | 1383 |
| 4 | 10.111 | 0.591 | 1089 |

Obtained results of measurement are listed in Table 4.1. The measured data and estimation values of read-time for one event data are plotted in Figure 4.2. As a result, the read-time for one event is reduced 4 times with 4 parallel readings: ~40msec down to ~10msec. In previous DAQ, read-time/event is about 20msec [32]. It means using new DAQ system with 4 parallel readings is 2 times faster than previous DAQ system.

**2. Evaluation with multiple event buffers**

Events occurrence in CANDLES follow random process. Interval distribution of these events is exponential distribution as following equation (section 4.5.5 in [28]):

$$P(t) = \mu e^{-\mu t} \tag{4.1}$$

where μ is the mean trigger rate. Statistical fluctuation of the random event occurrence causes dead time, while regular events do not cause dead time until the event processing rate exceed the readout rate. Figure 4.3 is an expression of event lost when using single event buffer. As we can see, even though read-time/event is reduced, there are still events lost. The event buffers help "de-randomize" the event processing and reduce events lost.

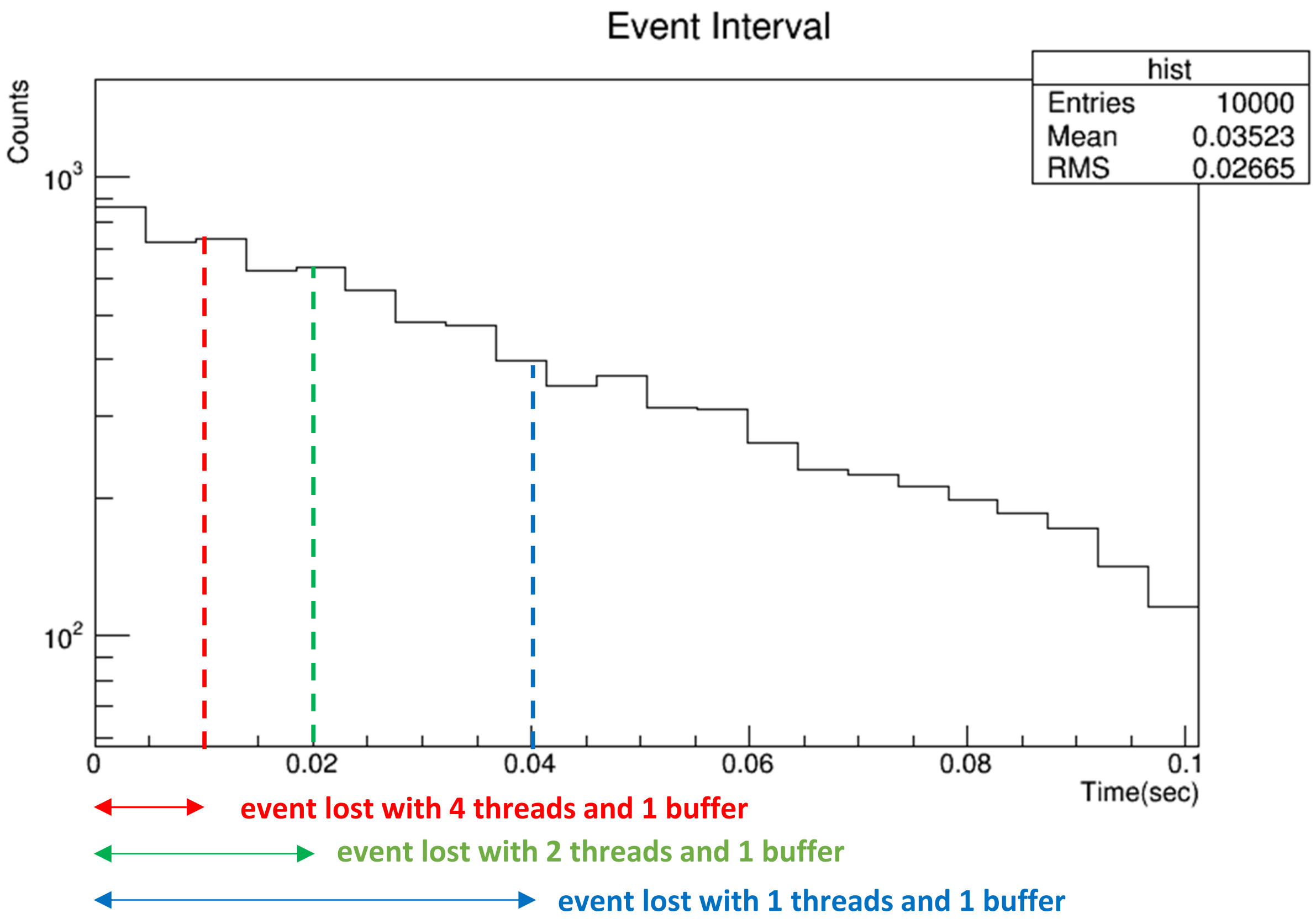

**Figure 4.4.** Expression of event lost when using single buffer. Histogram is the event interval distribution of 20cps trigger rate with zoom range (up to 0.1 sec). The data are taken after ~520 sec of measurement to obtain 10000 events.

With more event buffers, we can reduce more inefficiency. As mentioned in Chapter 3, DAQ inefficiency is proportional to inversed factorial of events buffers. In this test, we evaluate the influence of number of event buffers to inefficiency of new DAQ system. We trigger the DAQ system with random function generator. Output signals are generated with two frequencies ~40cps and ~100cps. These signals are fed to the input of one AMC-FADC. DAQ Middleware is configured with 4 parallel reading. We can set number of event buffers used in the test. Each input frequency from random function generator is tested with different number of event buffers (1 to 8).

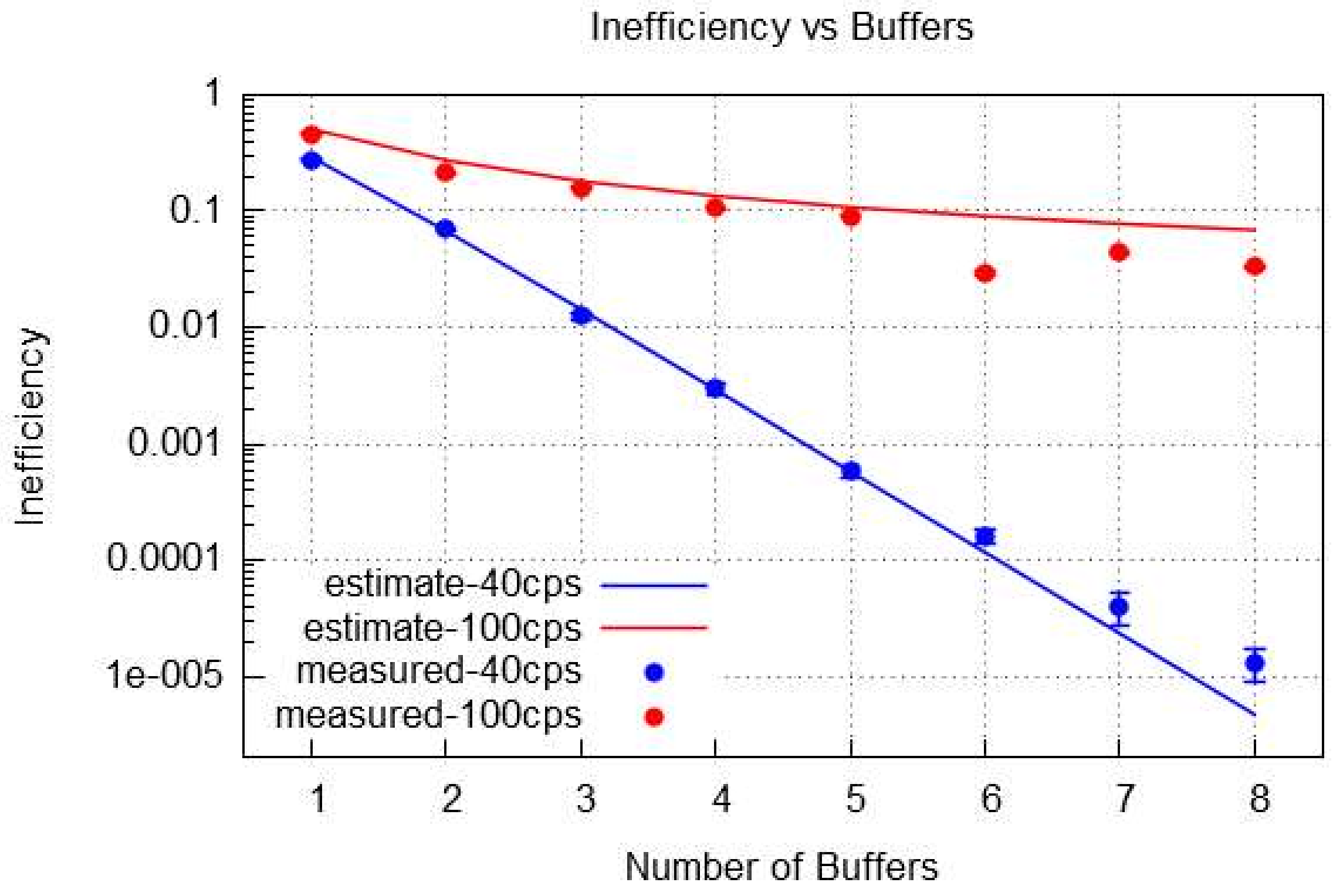

**Figure 4.5.** Inefficiency as a function of number of buffers. Solid lines are estimated inefficiency and circle points are measured inefficiency.

Number of trigger events and obtained events are counted at Master Module. They are namely Request and Accept, respectively. Readout packets from Master Module contains the count of Request and Accept. Inefficiency of DAQ system is computed by the ratio of lost events to trigger events as equation (4.3):

$$\text{Inefficiency} = \frac{\text{lost events}}{\text{total events}} = \frac{\text{Request} - \text{Accept}}{\text{Request}} \qquad (4.2)$$

Incoming events from function generator can only be obtained or lost. It is a dichotomous process, thus, we can apply binominal distribution to estimate the uncertainty (σ):

$$\sigma_{\text{Inefficiency}} = \frac{\sqrt{\text{N} \times (1 - \text{Inefficiency}) \times \text{Inefficiency}}}{\text{N}} \qquad (4.3)$$

where N is number of Request events generated by function generator.

**Table 4.2.** Obtained inefficiency corresponding to number of buffers

| **Trigger input** | **Buffer(s)** | **Run Time (sec)** | **Request Rate (cps)** | **Inefficiency (with uncertainty)** | **Inefficiency (estimated)** |
|---|---|---|---|---|---|
| 40cps | 1 | 171.63 | 37.995 | $2.72\times10^{-1} \pm 5.51\times10^{-3}$ | $2.88\times10^{-1}$ |
| | 2 | 371.44 | 38.582 | $6.94\times10^{-2} \pm 2.12\times10^{-3}$ | $6.69\times10^{-2}$ |
| | 3 | 304.69 | 38.856 | $1.25\times10^{-2} \pm 1.02\times10^{-3}$ | $1.42\times10^{-2}$ |
| | 4 | 607.54 | 38.134 | $2.94\times10^{-3} \pm 3.55\times10^{-4}$ | $2.88\times10^{-3}$ |
| | 5 | 3036.47 | 38.320 | $5.84\times10^{-4} \pm 7.08\times10^{-5}$ | $5.80\times10^{-4}$ |
| | 6 | 6404.68 | 37.408 | $1.63\times10^{-4} \pm 2.61\times10^{-5}$ | $1.17\times10^{-4}$ |
| | 7 | 6023.54 | 37.396 | $4.00\times10^{-5} \pm 1.33\times10^{-5}$ | $2.35\times10^{-5}$ |
| | 8 | 20727.00 | 37.240 | $1.30\times10^{-5} \pm 4.10\times10^{-6}$ | $4.72\times10^{-6}$ |
| 100cps | 1 | 413.40 | 94.519 | $4.62\times10^{-1} \pm 2.52\times10^{-3}$ | $5.02\times10^{-1}$ |
| | 2 | 187.71 | 94.827 | $2.11\times10^{-1} \pm 3.06\times10^{-3}$ | $2.72\times10^{-1}$ |
| | 3 | 186.45 | 94.814 | $1.58\times10^{-1} \pm 2.74\times10^{-3}$ | $1.80\times10^{-1}$ |
| | 4 | 211.49 | 94.014 | $1.05\times10^{-1} \pm 2.17\times10^{-3}$ | $1.35\times10^{-1}$ |
| | 5 | 182.19 | 93.112 | $9.00\times10^{-2} \pm 2.20\times10^{-3}$ | $1.08\times10^{-1}$ |
| | 6 | 158.74 | 93.020 | $2.91\times10^{-2} \pm 1.38\times10^{-3}$ | $9.02\times10^{-2}$ |
| | 7 | 384.26 | 93.692 | $4.40\times10^{-2} \pm 1.08\times10^{-3}$ | $7.78\times10^{-2}$ |
| | 8 | 511.54 | 93.408 | $3.28\times10^{-2} \pm 8.15\times10^{-4}$ | $6.85\times10^{-2}$ |

Results of inefficiency measurement are listed in Table 4.2. For comparison, estimation inefficiency at 40cps and 100cps is also computed with different number of event buffers. In figure 4.3, obtained data (circles) and estimation value (lines) are plotted.

Estimated values of inefficiency is computed with 40cps and 100cps. There are several data points far from estimation. It is caused by the fluctuation of input request rate since we use random function generator in our test. However, in general, measured data points are consistent with estimation. From observed results, we can conclude DAQ inefficiency is reduced with multiple event buffers.

**3. DAQ performance of new DAQ system**

In this section, we investigate DAQ performance with final configuration, which is using 4 parallel readings and 8 event buffers, at different trigger rate. The factor to evaluate performance is efficiency, which is calculated by ratio of obtained events to trigger events as equation (4.5):

$$\text{Efficiency} = \frac{\text{obtained events}}{\text{total events}} = \frac{\text{Accept}}{\text{Request}} \tag{4.4}$$

Since this is dichotomous process, we can still compute uncertainty of efficiency with binominal distribution:

$$\sigma_{\text{Efficiency}} = \frac{\sqrt{\text{N} \times (1 - \text{Efficiency}) \times \text{Efficiency}}}{\text{N}} \tag{4.5}$$

As abovementioned, Request and Accept are contained in Master Module data packet. In this test, signals generated from random function generator are fed to AMC-FADC for triggering. Frequency of these signals are varied from ~20cps to ~100cps. In order to see the influence of event buffers, efficiency is tested with not only 8 buffers but also 3 buffers. Obtained results of new DAQ system are described in Table 4.3.

Data of previous DAQ system are also plot to see the improvement of new DAQ system. These data are obtained at [29]. Previous DAQ system uses three event buffers distributed in three PCs (1 buffer/PC). Data in [32] were obtained with two of the third (2/3) of FADC channels in previous CANDLES set up. Since the number of FADC channels is proportional to read-time/event, it means the read-time/event for full set of previous DAQ system is 1.5 times longer.

**Table 4.3.** Efficiency of new DAQ system using 4 parallel readings and 8 event buffers

| Buffer(s) | Run Time (sec) | Request Rate (Hz) | Efficiency (with uncertainty) | Efficiency (estimated) |
|---|---|---|---|---|
| 3 | 789.02 | 18.662 | 99.8642% ± 0.0304% | 99.8378% |
| | 304.69 | 38.856 | 98.7499% ± 0.1021% | 98.5817% |
| | 262.03 | 59.997 | 93.2892% ± 0.1996% | 95.1861% |
| | 209.89 | 75.416 | 91.5219% ± 0.2214% | 89.4026% |
| | 186.45 | 94.814 | 84.2460% ± 0.2740% | 81.9772% |
| 8 | 228175.00 | 18.618 | 100.0000% ± 0.0000% | 100.0000% |
| | 20727.00 | 37.240 | 99.9987% ± 0.0004% | 99.9995% |
| | 1318.42 | 59.442 | 99.8009% ± 0.0159% | 99.9569% |
| | 183.30 | 74.714 | 99.2187% ± 0.0752% | 99.0275% |
| | 511.54 | 93.408 | 96.7205% ± 0.0815% | 93.1506% |

**Table 4.4.** Efficiency of previous DAQ system. Table consists of data (of request rate and accept rate) extracted from [32] and data converted to full set of FADC channels. Only data with converted request rate from ~20cps to ~100cps are mentioned in this table.

| *Data from [32]* | | *Data converted* | | |
|---|---|---|---|---|
| Request Rate (Hz) | Accepted Rate (Hz) | Request Rate (Hz) | Accepted Rate (Hz) | Efficiency (with uncertainty) |
| 30.586 | 30.116 | 20.390 | 20.078 | 98.47% ± 2.22% |
| 59.580 | 56.203 | 39.720 | 37.469 | 94.33% ± 3.00% |
| 120.977 | 95.230 | 80.651 | 63.487 | 78.72% ± 3.72% |

As discussion in Chapter 3, inefficiency or efficiency depends on multiplication of trigger rate and read-time/event. Thus, with a same efficiency, we have the conversion:

$$(\mu T)_{test} = (\mu T)_{real} \tag{4.6}$$

where $(\mu T)_{test}$ and $(\mu T)_{real}$ are respectively multiplication in the test of [32] and real set up of previous DAQ system. Hence, we can deduce trigger rate in real set up:

$$\Rightarrow \mu_{real} = \mu_{test} \times \left( \frac{T_{test}}{T_{real}} \right) = \frac{2}{3} \mu_{test} \tag{4.7}$$

Original data and converted data of previous DAQ system are expressed in Table 4.4. Uncertain of efficiency of previous DAQ system is also computed with binominal distribution. All data of new and old DAQ system are plotted in Figure 4.4.

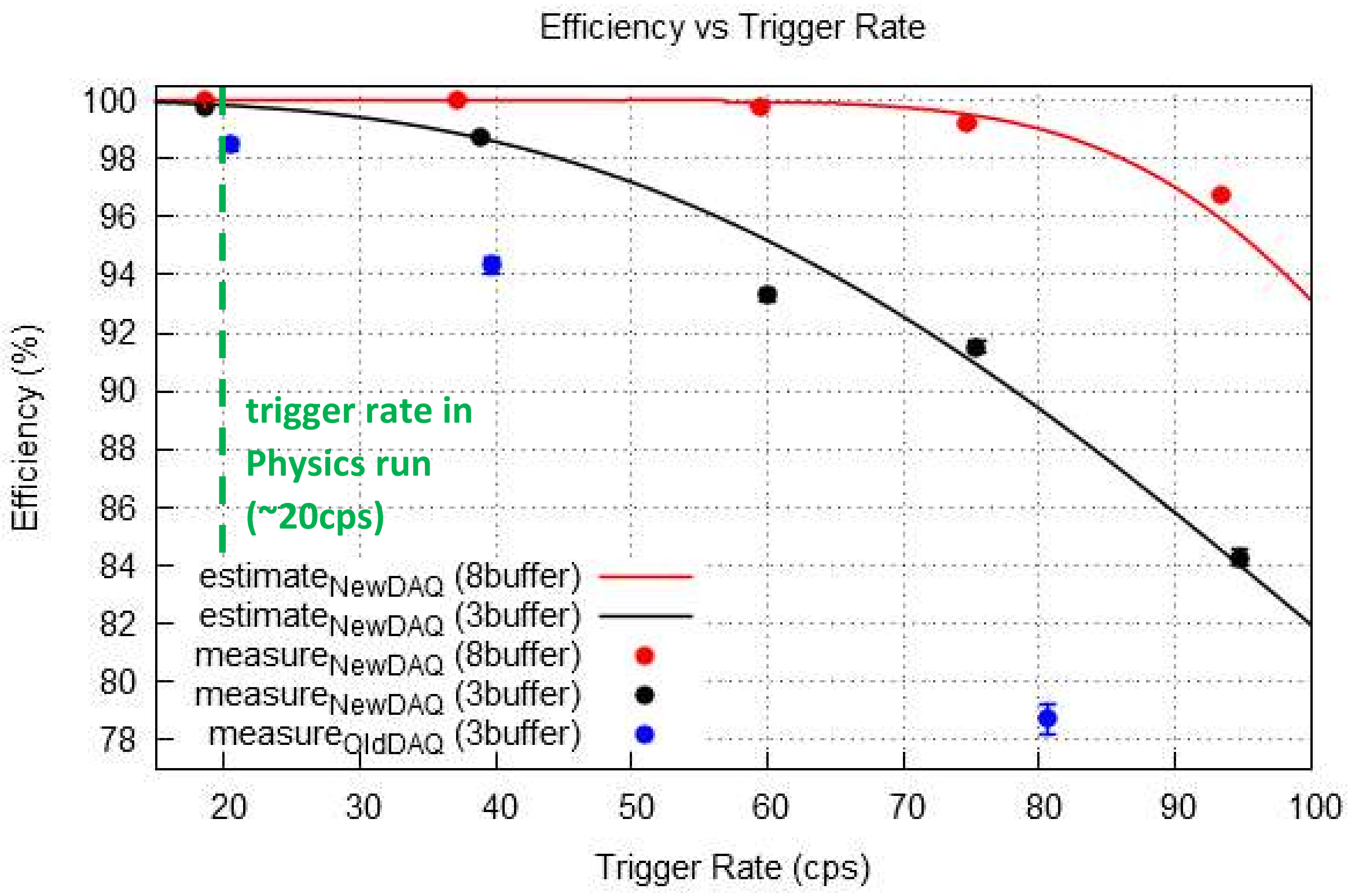

**Figure 4.6.** Efficiency measured with various random trigger rate. Solid lines are estimated inefficiency. Data new of DAQ system with 8 event buffers (red point) and 3 event buffers (black point) are plotted to compare with previous DAQ system (blue point).

Estimation lines of new DAQ system are plotted in Figure 4.4 while estimation of previous DAQ is not. The reason is the read-time/event of previous DAQ system is extended at trigger rate higher than 20cps [32]. According to obtained data, there are several comments:

- We achieve higher efficiency of new DAQ system with 8 event buffers compared to efficiency of new DAQ with 3 buffers due to the influence of event buffers to efficiency.
- Comparing efficiency with 3 buffers of new DAQ and previous DAQ, efficiency of new DAQ is higher than previous DAQ's at trigger rate higher or equal to 20cps. It is influenced by read-time/event: read-time/event of new DAQ in this test is ~10msec while read-time/event of previous DAQ is ~20msec at 20cps and being extended at higher trigger rate.
- Trigger rate in Physics run of CANDLES experiment is 20cps [32]. Thus, efficiency at this trigger rate is very important to consider. At this trigger rate, with 4 parallel readings and 8 buffers, there is no event lost after ~63 hours of data taking (corresponding to 4,248,248 events). With this data, we can evaluate the lower limit of efficiency as well as upper limit of inefficiency with corresponding Confidence Level (CL) and number of trigger events (N) by following equation (page 100 in [28]):

$$\varepsilon_O = (1 - CL)^{1/(N+1)} = 1 - \overline{\varepsilon_O} \tag{4.8}$$

where $\varepsilon_O$ is lower limit of efficiency and $\overline{\varepsilon_O}$ is the upper limit of inefficiency. With obtained data with new DAQ system using 4 parallel reading and 8 event buffers, upper limit of DAQ inefficiency with 99% of CL is $1.084 \times 10^{-6}$. This means efficiency is very close to 100%. On the other hand, efficiency of previous DAQ system is around 98% to 99%. According to obtained data, it can be concluded, at physics run, efficiency of new DAQ system is higher than previous one. Improved efficiency with new DAQ is enough performance for CANDLES experiment.

Tagging efficiency in CANDLES experiment depends on efficiency of data taking and analysis software. At the current status, tagging efficiency of $^{208}$Tl is about 60% [29] with previous DAQ system. The new DAQ system with nearly 100% efficiency has a small influence on the tagging efficiency. For higher tagging efficiency, it needs improvements analysis software (and also passive shielding). The improvement of DAQ system in this research will be necessary in the future when the analysis software is updated.

# SUMMARY

One of the serious background of CANDLES comes from beta decay of $^{208}Tl$. To remove the beta decay of $^{208}Tl$, we tag the preceding alpha decay of $^{212}Bi$. This tagging method requires minimized dead-time of data taking. A new DAQ system was introduced in CANDLES experiment using Flash-ADCs with 8-buffers and SpW-GbE network for data readout. To reduce the dead-time, we realized our new DAQ system with 4 parallel reading processes and used 8 event buffers. As a result, we achieve no event lost after 63 hours of data taking at 20 cps, which is CANDLES trigger rate. It is in equivalent to an upper limit $1.084\times10^{-6}$ (C.L. 99%) of inefficiency. The new DAQ system has enough performance for CANDLES experiment. Tagging efficiency depends on high DAQ efficiency and good data analysis. Because we achieve very high efficiency, tagging efficiency in CANDLES mainly depends on software analysis.

# FUTURE WORK

Tagging efficiency of CANDLES depends on efficiency of data taking and offline software analysis. Now, efficiency of data taking is really high. To remove background comes from sequential decays, such like $^{208}$Tl, we need to improve analysis software. My current work is one of the tasks to achieve background-free condition. In the current status, three sequential decays with half-lives less than one second were used for impurity determination assuming secular equilibrium. However, there are other sequential decays which can be used for impurities study. My research work will focus on using more sequential decays for determination and confirming radioactive impurity amount in CANDLES. Additionally, the current status uses a set of beta-decay and alpha-decay for evaluating contamination. However, beta-decay has fluctuated kinetic energy while alpha-decay has discrete released energy. Using a set of alpha-decay and alpha-decay may give better evaluation of contamination. These decays can be found in other sequential decays. In my future work, I will also try to realize an analysis software using these alpha decays.

# ACKNOWLEDGEMENT

This thesis cannot be completed without support of many people:

I would like to give my appreciation to Prof. Masaharu Nomachi. He accepted me as a Master student in his research group. He provides me the Physics and technical stuffs. With his help, the Master thesis is much improved. I really appreciate having studied under his supervision.

I would like to thank Ass. Prof. Yorihihito Sugaya for comments and discussions in this research work. He gave me many advices during measurement.

I would like to thank Mr. Tsuyoshi Maeda for tutoring in DAQ, Linux, CANDLES and also Japanese daily life. I would like to thank Mr. Kazuki Kanagawa and Mr. Masahito Tuzuki for cooperation in DAQ development. Thank you all of you for not only being my lab-mates, but also good friends, good collaborators.

Thanks my teachers in VNUHCM University of Science for supporting me the fundamental knowledge in Nuclear Physics and experiments. Especially, I appreciate Prof. Chau Van Tao Dr. Vo Hong Hai for supporting me to study in Osaka University.

Thanks my family for always supporting, helping and understanding. You help me overcome many difficulties in this oversea life. Thanks to my aunt for her understanding and encouragement. Thanks my love for your encouragement and forcing me to this Physics life. I join this "new life" without any regret.

Bui Tuan Khai.

# REFERENCES

[1] G. J. Neary, "The β-ray spectrum of radium E", Proceedings of the Royal Society of London, Series A, Mathematical and Physical Science (1939), pp.71-87.

[2] Samuel. S. M. Wong, "Physics Textbook: Introductory Nuclear Physics – Second Edition", Wiley-VCH published.

[3] Andrea Giuliani and Alfredo Poves, "Review Article: Neutrinoless Double-Beta Decay", Advances in High Energy Physics, Vol. 2012, Article ID 857016, doi:10.1155/2012/857016.

[4] A. Balysh et al., Double Beta Decay of 48Ca, Phys. Rev. Lett. 77 (Dec, 1996) 5186.

[5] A. Bakalyarov et al., "Search for β− and β−β− decays of 48Ca", Nuc. Phys. A., Vol. 700 (2002), pp.17-24.

[6] M. Aunola, J. Suhonen, T. Siiskonen, "Shell-model study of the highly forbidden beta decay 48Ca → 48Sc", Europhys. Lett. 46 (1999) 577.

[7] Ruben Saakyan, "Two-Neutrino Double-Beta Decay", Annu. Rev. Nucl. Part. Sci. 2013, doi: 10.1146/annurev-nucl-102711-094904.

[8] GERDA Collaboration Collaboration, M. Agostini et al., Measurement of the half-life of the two-neutrino double beta decay of Ge-76 with the Gerda experiment, J. Phys. G 40 (2013) 035110, arXiv:1212.3210 [nucl-ex].

[9] NEMO-3 Collaboration Collaboration, J. Argyriades et al., Measurement of the two neutrino double beta decay half-life of 96Zr with the NEMO-3 detector, Nucl. Phys. A 847 (2010) 168, arXiv: 0906.2694 [nucl-ex].

[10] NEMO Collaboration Collaboration, R. Arnold et al., First results of the search of neutrinoless double beta decay with the NEMO 3 detector, Phys. Rev. Lett. 95 (2005) 182302, arXiv:hep-ex/0507083 [hep-ex].

[11] NEMO Collaboration Collaboration, L. Simard et al., The NEMO-3 results after completion of data taking, Journal of Physics: Conference Series 375 (2012) no. 4, 042011.

[12] NEMO-3 Collaboration Collaboration, R. Arnold et al., Measurement of the Double Beta Decay Half-life of 130Te with the NEMO-3 Detector, Phys. Rev. Lett. 107 (2011) 062504, arXiv: 1104.3716 [nucl-ex].

[13] [Phys.Rev.C.Vol.89] J. B. Albert et al. (EXO Collaboration), "Improved measurement of the 2νββ half-life of $^{136}$Xe with the EXO-200 detector", Phys. Rev. C 89, 015502 (2014).

[14] [Phys.Rev.C.Vol.80] J. Argyriades et al. (NEMO Collaboration), "Measurement of the double-β decay half-life of $^{150}$Nd and search for neutrinoless decay modes with the NEMO-3 detector", Phys. Rev. C 80, 032501 (2009).

[15] Andrea Giuliani and Alfredo Poves, "Neutrinoless Double-Beta Decay", Advances in High Energy Physics Volume 2012, Article ID 857016, doi:10.1155/2012/857016.

[16] S. Umehara et al., "Neutrino-less double-β decay of $^{48}$Ca studied by $CaF_2$(Eu) scintillators", Phys. Rev. C, Vol. 78, 058501, 2008.

[17][Phys.Rev.Lett.Vol.111] M. Agostini et al. (GERDA Collaboration), "Results on Neutrinoless Double-β Decay of $^{76}$Ge from Phase I of the GERDA Experiment", Phys. Rev. Lett. 111, 122503 (2013).

[18] [Thesis.UCL.2013] J. Mott, "Search for double beta decay of 82Se with the NEMO-3 detector and development of apparatus for low-level radon measurements for the SuperNEMO experiment", PhD thesis, University College London, 2013.

[19] [Phys.Rev.D.Vol.89] R. Arnold et al. (NEMO-3 Collaboration), "Search for neutrinoless double-beta decay of $^{100}$Mo with the NEMO-3 detector", Phys. Rev. D 89, 111101 (2014).

[20] [Phys.Rev.C.Vol.68] F. A. Danevich et al., "Search for 2β decay of cadmium and tungsten isotopes: Final results of the Solotvina experiment", Phys. Rev. C 68 (Sep, 2003) 035501.

[21] [Ast.Phys.Vol.34] E. Andreotti et al., "$^{130}$Te Neutrinoless Double-Beta Decay with CUORICINO", Astropart. Phys. 34 (2011) 822, arXiv:1012.3266 [nucl-ex].

[22] [Phys.Rev.Lett.Vol.110] KamLAND-Zen Collaboration Collaboration, A. Gando et al., "Limit on Neutrinoless ββ Decay of Xe-136 from the First Phase of KamLAND-Zen and Comparison with the Positive Claim in Ge-76", Phys. Rev. Lett. 110 (2013) no. 6, 062502, arXiv:1211.3863 [hep-ex].

[23] Wikipedia https://en.wikipedia.org/

[24] Hidekazu Kakubata, "Study of Backgrounds in CANDLES to Search for Double Beta Decays of 48Ca", Dissertation in Physics, Osaka University, (2015).

[25] Y.Hirano, Doctoral thesis, Osaka University (2008).

[26] S. Yoshida et al., "Ultra-violet wavelength shift for undoped CaF2 scintillation detector by two phase of liquid scintillator system in CANDLES", Nucl. Instr. and Meth. A 601 (2009) 282-293.

[27] Kishomoto Lab homepage: http://wwwkm.phys.sci.osaka-u.ac.jp/

[28] W. R. Leo, "Techniques for Nuclear and Particle Physics Experiments – Second Revised Edition", ISBN 3-540-57280-5.

[29] T Iida, K Nakajima, et al., "Status and future prospect of 48Ca double beta decay search in CANDLES", Journal of Physics: Conference Series 718 (2016) 062026.

[30] S. Umehara et al., "Search for Neutrino-less Double Beta Decay with CANDLES", Physics Procedia, Vol.61 (2015), pp.283-288.

[31] S. Umehara, "Study of Double Beta Decays of $^{48}$Ca with $CaF_2$ Scintillators", Dissertation in Physics, Osaka University, (2004).

[32] Suzuki et al., "New DAQ System for the CANDLES Experiment", IEEE Trans. Nucl. Sci., Vol. 62, No. 3, pp. 1122-1127, Jun. 2015.

[33] PICMG homepage: http://www.picmg.org

[34] VadaTech Inc., “MicroTCA Overview”, online document, 2014.

[35] Tsuyoshi Maeda, “CANDLESによる二重ベータ崩壊の研究 (101): Micro-TCA規格の 500MHz-FADC の開発” (Presentation), The JPS 70th Annual Meeting, Waseda University, Japan (March 2015).

[36] Tsuyoshi Maeda, “New Micro-TCA base DAQ system for CANDLES” (Poster), International Symposium on revealing the history of the universe with underground particle and nuclear research 2016, University of Tokyo, Tokyo, Japan.

[37] T. Maeda et al., “The CANDLES Trigger System for the Study of Double Beta Decay of $^{48}Ca$”, IEEE Trans. Nucl. Sci., Vol. 62, No. 3, pp. 1128-1134, Jun. 2015.

[38] Star Dundee homepage: https://www.star-dundee.com/

[39] Shimafuji homepage: http://www.shimafuji.co.jp/

[40] Takayuki Yuasa (JAXA), “SpaceWire RMAP Library v2: User Guide”, Jan. 2012, online: https://github.com/yuasatakayuki/SpaceWireRMAPLibrary

[41] Steve Parkes and Chris McClements, “SpaceWire Remote Memory Access Protocol”, online: http://spacewire.esa.int/content/TechPapers/TechPapers.php

[42] E. Inoue et al., “A DAQ System for CAMAC Controller CC/NET Using DAQ-Middleware”, Journal of Physics: Conference Series 219 (2010) 022036, doi:10.1088/1742-6596/219/2/022036.

[43] Agilent Technologies, “Agilent Technologies InfiniiVision 7000B Series Oscilloscopes – Data Sheet”, http://www.farnell.com/datasheets/1749653.pdf

[44] P. Cerello et al., “The FINUDA data acquisition system a C/C++ DAQ based on ROOT as event monitor”, IEEE Transactions on Nuclear Science, Vol.45, No.4, pp.1973-1977 (1998).

# Appendix A

| | | | | | | | | | | | | | | |
|---|---|---|---|---|---|---|---|---|---|---|---|---|---|---|
| 2015/11/21：初版作製<br>2016/2/21 FadcID追加 | | | | | | | | | | | | | | |
| MTCA | MCH-1 | AMC-01 | AMC-02 | AMC-03 | AMC-04 | AMC-05 | AMC-06 | AMC-07 | AMC-08 | AMC-09 | AMC-10 | AMC-11 | AMC-12 | MCH-2 |
| MTCA-1 | MCH01 | FADC01 | FADC02 | FADC03 | FADC04 | | FADC05 | | FADC06 | FADC07 | FADC08 | FADC09 | | |
| Module-ID | 23015002 | 24033001 | 24077002 | 24077003 | 24077004 | | 24077005 | | 24077006 | 24077007 | 24077008 | 24077009 | | |
| Fadc-ID | mch00 | fadc01<br>fadc00 | fadc03<br>fadc02 | fadc05<br>fadc04 | fadc07<br>fadc06 | | fadc09<br>fadc08 | | fadc11<br>fadc10 | fadc13<br>fadc12 | fadc15<br>fadc14 | fadc17<br>fadc16 | | |
| PMT-ID | | PMT02<br>PMT01 | PMT04<br>PMT03 | PMT06<br>PMT05 | PMT08<br>PMT07 | | PMT10<br>PMT09 | | PMT12<br>PMT11 | PMT14<br>PMT13 | 62SUM<br>PMT15 | 62SUM<br>PMT16 | | |
| | | | | | | | | | | | | | | |
| MTCA | MCH-1 | AMC-01 | AMC-02 | AMC-03 | AMC-04 | AMC-05 | AMC-06 | AMC-07 | AMC-08 | AMC-09 | AMC-10 | AMC-11 | AMC-12 | MCH-2 |
| MTCA-2 | MCH02 | FADC10 | FADC11 | FADC12 | FADC13 | | FADC14 | | FADC15 | FADC16 | FADC17 | FADC18 | | |
| Module-ID | 24104001 | 24077010 | 24077011 | 24077012 | 24077013 | | 24077014 | | 25029001 | 25029002 | 25029003 | 25029004 | | |
| Fadc-ID | mch01 | fadc01<br>fadc00 | fadc03<br>fadc02 | fadc05<br>fadc04 | fadc07<br>fadc06 | | fadc09<br>fadc08 | | fadc11<br>fadc10 | fadc13<br>fadc12 | fadc15<br>fadc14 | fadc17<br>fadc16 | | |
| PMT-ID | LowDG | PMT18<br>PMT17 | PMT20<br>PMT19 | PMT22<br>PMT21 | PMT24<br>PMT23 | | PMT26<br>PMT25 | | PMT28<br>PMT27 | PMT30<br>PMT29 | 62SUM<br>PMT31 | 62SUM<br>PMT32 | | |
| | | | | | | | | | | | | | | |
| MTCA | MCH-1 | AMC-01 | AMC-02 | AMC-03 | AMC-04 | AMC-05 | AMC-06 | AMC-07 | AMC-08 | AMC-09 | AMC-10 | AMC-11 | AMC-12 | MCH-2 |
| MTCA-3 | MCH03 | FADC19 | FADC20 | FADC21 | FADC22 | | FADC23 | | FADC24 | FADC25 | FADC26 | FADC27 | | |
| Module-ID | 24104002 | 25029005 | 25029006 | 25029007 | 25029008 | | 25029009 | | 25029010 | 25029011 | 25029012 | 25029013 | | |
| Fadc-ID | mch02 | fadc01<br>fadc00 | fadc03<br>fadc02 | fadc05<br>fadc04 | fadc07<br>fadc06 | | fadc09<br>fadc08 | | fadc11<br>fadc10 | fadc13<br>fadc12 | fadc15<br>fadc14 | fadc17<br>fadc16 | | |
| PMT-ID | MB<br>UpperVeto | PMT34<br>PMT33 | PMT36<br>PMT35 | PMT38<br>PMT37 | PMT40<br>PMT39 | | PMT42<br>PMT41 | | PMT44<br>PMT43 | PMT46<br>PMT45 | 62SUM<br>PMT47 | 6SUM<br>PMT48 | | |
| | | | | | | | | | | | | | | |
| MTCA | MCH-1 | AMC-01 | AMC-02 | AMC-03 | AMC-04 | AMC-05 | AMC-06 | AMC-07 | AMC-08 | AMC-09 | AMC-10 | AMC-11 | AMC-12 | MCH-2 |
| MTCA-4 | MCH04 | FADC28 | FADC29 | FADC30 | FADC31 | | FADC32 | | FADC33 | FADC34 | FADC35 | FADC36 | FADC37 | |
| Module-ID | 25029001 | 25029014 | 25029015 | 25029016 | 25029017 | | 25029018 | | 25029019 | 25029020 | 25029022 | 25029023 | 25029024 | |
| Fadc-ID | mch03 | fadc01<br>fadc00 | fadc03<br>fadc02 | fadc05<br>fadc04 | fadc07<br>fadc06 | | fadc09<br>fadc08 | | fadc11<br>fadc10 | fadc13<br>fadc12 | fadc15<br>fadc14 | fadc17<br>fadc16 | fadc19<br>fadc18 | |
| PMT-ID | Cosmic | PMT50<br>PMT49 | PMT52<br>PMT51 | PMT54<br>PMT53 | PMT56<br>PMT55 | | PMT58<br>PMT57 | | PMT60<br>PMT59 | PMT62<br>PMT61 | 6SUM<br>(G:1/200)<br>Reserve | 3SUM<br>(G:1/200)<br>+3SUM<br>(G:1/200) | 3SUM<br>(G:1/25)<br>+3SUM<br>(G:1/25) | |

**Figure A.** PMT-to-FADC mapping in new DAQ system of CANDLES

# Appendix B

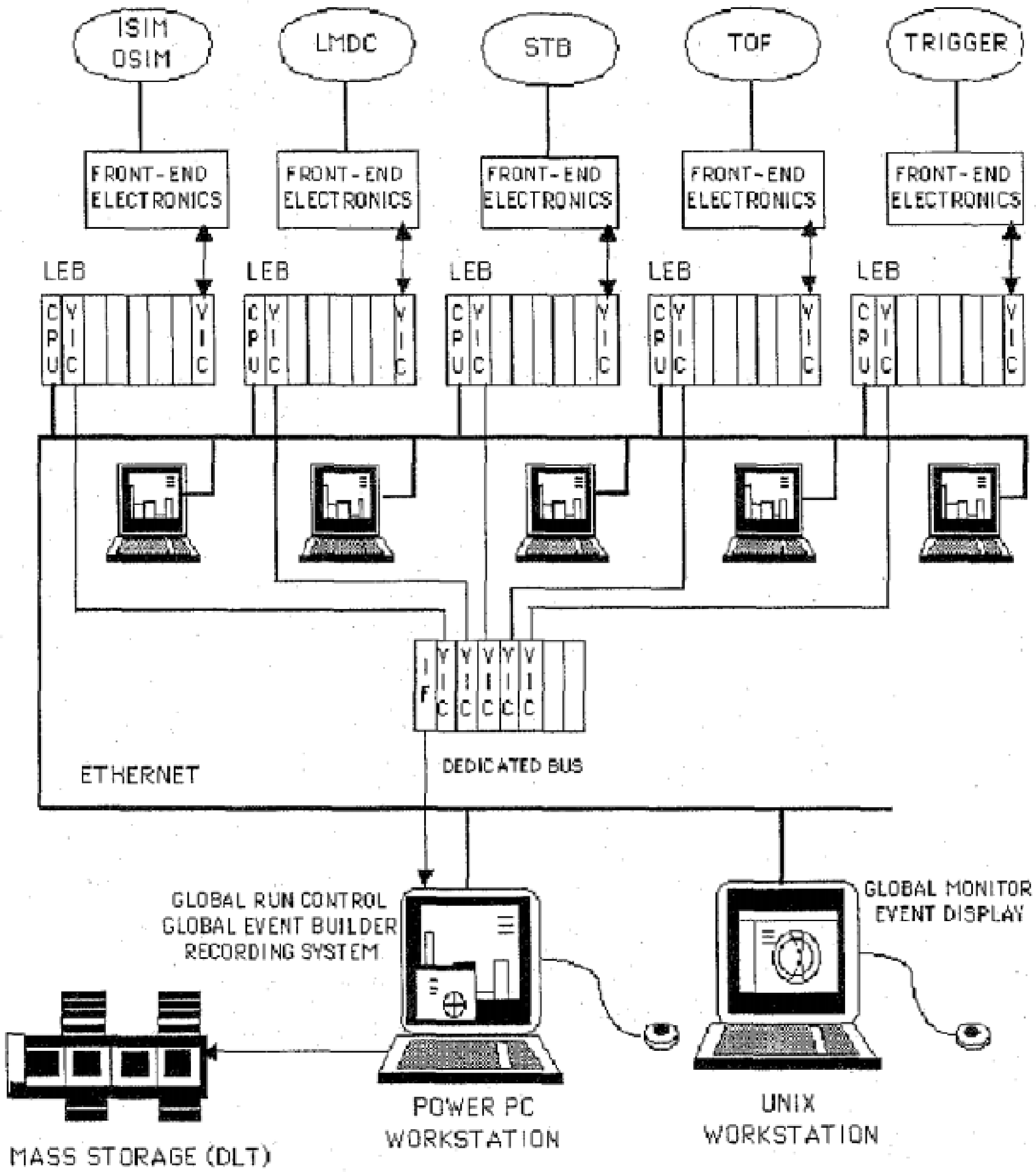

**Figure B.** Schematic architecture of FINUDA DAQ system [44]. Slave PCs transfer data to Power PC after collecting data from detectors. Power PC is in charge of run control, event builder and recording system.